\documentclass[12pt]{article}
\usepackage{etex}
\usepackage{fullpage}
\usepackage{setspace}
\usepackage{comment}
\usepackage[pdftex]{graphicx}
\usepackage{rotating}
\usepackage{amsmath}
\usepackage{mdframed}
\usepackage{amsthm}
\usepackage{amssymb}
\usepackage{graphicx}
\usepackage{fullpage}
\usepackage{setspace}
\usepackage{natbib}
\usepackage{color}
\usepackage{booktabs}
\usepackage{caption}
\usepackage[linesnumbered, ruled]{algorithm2e}
\SetKwRepeat{Do}{do}{while}%

\usepackage{tikz}
\usepackage{pgf}
\usepackage{pgfplots}
\pgfplotsset{compat=1.16}
\usepackage{textpos}
\usepackage{graphicx}
\usepackage{pstricks}
\usepackage{framed}
\usepackage{varwidth}
\usepackage[normalem]{ulem}
\usepackage{xstring}
\usetikzlibrary{shapes}
\usetikzlibrary{arrows,decorations.pathmorphing,backgrounds,fit,
positioning,shapes.symbols,chains}

\newcommand{\T}{\intercal}

\newcommand{\pn}{\mathbb{P}_{n}}
\newcommand{\piOpt}{\pmb{\pi}^{\mathrm{opt}}}
\newcommand{\lamOpt}{\pmb{\lambda}^{\mathrm{opt}}}

\newcommand{\OMIT}[1]{\relax}   
\def\text{{\rm}}

 \newcommand{\bma}[1]{\mbox{\boldmath $#1$}}

 \newcommand{\bA}{ {\mathbf{A}} }

 \newcommand{\bc}{ {\mathbf{c}} }

 \newcommand{\bH}{ {\mathbf{H}} }
 \newcommand{\bh}{ {\mathbf{h}} }

 \newcommand{\bQ}{ {\mathbf{Q}} }

 \newcommand{\bV}{ {\mathbf{V}} }

 \newcommand{\bX}{ {\mathbf{X}} }
 
 \newcommand{\bY}{ {\mathbf{Y}} }

 \newcommand{\bepsilon}{ {\bma{\epsilon}} }

 \newcommand{\blambda}{ {\pmb{\lambda}} }
 \newcommand{\bLambda}{ {\pmb{\Lambda}} }
 
 \newcommand{\bdelta}{ {\pmb{\delta}} }

 \newcommand{\bSigma}{ {\pmb{\Sigma}} }
 \newcommand{\bpi}{\pmb{\pi}}
 \newcommand{\bPi}{\pmb{\Pi}}

\newtheorem{thm}{Theorem}[section]
\newtheorem{lem}[thm]{Lemma}
\newtheorem{cor}[thm]{Corollary}
\theoremstyle{definition}

\newtheorem{rmrk}[thm]{Remark}

\begin{document}
\begin{center}
\textbf{Optimal treatment strategies for prioritized outcomes\\ 
Kyle Duke, Eric B. Laber, Marie Davidian, Michael Newcomb, Brian Mustanski}
\end{center}
\begin{abstract}
Dynamic treatment regimes formalize precision medicine as a sequence
of decision rules, one for each stage of clinical intervention, that map
current patient information to a recommended intervention.  Optimal regimes
are typically defined as maximizing some functional of a scalar outcome's 
distribution, e.g., 
the distribution's mean or median.  However, in many clinical applications, 
there are multiple outcomes of interest.  We consider the problem of estimating 
an optimal regime when there
are multiple outcomes that are ordered by priority but which cannot be readily 
combined by domain experts into a meaningful single scalar outcome.  
We propose a definition of
optimality in this setting and show that an optimal regime with respect to 
this definition leads to maximal
mean utility under a large class of utility functions.  Furthermore, we use
inverse reinforcement learning to identify a composite outcome that most
closely aligns with our definition within a pre-specified class. Simulation 
experiments and an application to data from a sequential
multiple assignment randomized trial (SMART) on HIV/STI 
prevention illustrate the usefulness of the
proposed approach.  
\end{abstract}

\section{Introduction}
Clinical interventions often target multiple components
of a patient's health and behavior.  For example, this 
work is motivated by our involvement in a
sequential multiple assignment randomized
trial \citep[SMART][]{lavori2004dynamic,murphy2005experimental}
on human immunodeficiency virus (HIV)/sexually transmitted infection (STI) 
prevention through electronic health (eHealth) interventions 
in adolescent men 
who have sex with men \citep[][]{mustanski2020evaluation}.  
In this study, there are three primary outcomes of 
interest which are targeted by the interventions under study: 
(i) risky behavior, e.g., condomless
anal sex; (ii) regular HIV/STI testing; and (iii) 
self-efficacy for condom use.  These outcomes are ordered 
by importance, i.e., they are {\em prioritized} 
\citep[][]{finkelstein1999combining,buyse2010generalized}, 
but it is not clear how one should define
an optimal intervention strategy that reflects
these priorities.

One approach is to combine the outcomes into a single
composite outcome \citep[][]{thall2020statistical}.  For example, one 
might center and scale
the outcomes and construct a convex combination 
with larger weights on higher priority outcomes.  However,
in many settings, it is difficult for domain experts 
to construct such 
a composite outcome and nearly impossible to 
understand {\em a priori} how the choice of composite
outcome will affect an estimated intervention 
strategy 
\cite[see][
for additional discussion and references on 
the challenges associated with composite outcome elicitation]{lizotte2012linear,laber2014set,lizotte2016multi}.  

In the context of prioritized composite outcomes,
especially with time-to-event data, a common composite is the 
so-called win ratio 
\citep[][]{finkelstein1999combining,pocock2012win,oakes2016win,mao2021class}. 
Broadly, the win ratio is based on the sequential comparison of 
outcomes between two patients, say in treatment
in control, in order of priority.  First, the patients are compared on 
the highest priority outcome, if no winner
can be determined (say because the outcomes are missing due to censoring), the 
patients are compared in terms of the 
next highest priority outcome and so on until a winner is found
or the outcomes are exhausted in which case a tie is declared. The win ratio is the 
probability that a randomly selected patient from the treatment 
group will `win' against a randomly selected patient from the
control group.  
The win ratio has heretofore only been developed for comparisons of fixed 
treatments in a single time point setting. For completeness, we derive
a generalization of the win ratio to the comparison of multi-stage
treatment regimes and a partial order in which outcomes are `comparable'
if they are clinically meaningfully different.  However, in addition to the
criticisms of composite outcomes raised above, it is not straightforward 
to define an optimal regime using the win ratio as it need not be transitive
across regimes.
Furthermore, even if one chooses a suitable tie-breaking scheme, and thereby
fixes some notion of optimality using the win ratio, the result obscures the
importance of each outcome and is difficult to interpret.

Another closely related approach is to optimize one primary outcome subject to a constraint
on one or more secondary outcomes, e.g., maximize efficacy 
subject to a constraint on harm
\citep[][]{linn2015chapter,laber2018identifying,wang2020learning,illenberger2021identifying}.
Selecting the constraints is
equivalent (albeit implicitly) to choosing a composite outcome; thus, unless there
is strong clinical evidence to inform the constraints, this approach is subject
to many of the pitfalls associated with outcome elicitation.  
Furthermore, constrained optimization does not always align with clinical
objectives; e.g., in our motivating example, a constrained approach
would suggest minimizing risky behaviors
subject to constraints on testing and self-efficacy, which 
corresponds to a clinically unrealistic trade-off in which
unsafe behavior with sufficiently high self-efficacy is preferred
to no unsafe behavior.

Rather than trying to construct a composite outcome through 
direct elicitation, one can attempt to estimate a composite
using item response theory  \citep[IRT][]{embretson2013item}.
IRT has been used to estimate individual patient
preferences and subsequently estimate optimal
intervention strategies based on these preferences 
\citep[][]{butler2018incorporating}. However, the application of these
methods requires a high-quality instrument to measure
outcome preferences.  Unfortunately, such instruments
are not available for many applications including the
one we consider here.

Our approach is based on defining an optimal regime in a way
that directly aligns with clinical objectives. 
We then show
that the optimal regime under this definition maximizes expected
utility over a large class of (essentially all) utility
functions that are consistent with the prioritization over
outcomes.  We propose an estimator of this optimal regime and
derive asymptotic approximations to the sampling distribution
of the marginal mean outcome under the estimated regime.  
We also use inverse reinforcement learning \citep[][]{arora2021survey}
to estimate a composite outcome which the estimated optimal
regime approximately maximizes. This estimated composite can
generate new insights about the relative importance of each 
outcome in the optimal regime.

In Section \ref{sec:setup}, we set notation, present
our definition of 
an optimal regime under prioritized outcomes, and show that
it maximizes utility under a class of priority-respecting
utility functions.   
In Section \ref{sec:Estimation}, we describe estimators
of the optimal regime.  
In Section \ref{sec:Theory},
we derive large sample inference procedures for the
proposed estimators.  In Section \ref{sec:Sims},
we examine finite sample properties of the propsed estimation
and inference procedures via simulation experiments. 
We apply the proposed methods to the HIV/STI prevention
study in Section \ref{sec:CaseStudy}.

\section{Setup and notation}\label{sec:setup}
We assume data are collected in a $K$-stage 
sequential multiple assignment randomized trial 
\citep[SMART][]{lavori2000design,lavori2004dynamic,murphy2005experimental} conducted with 
$n$ patients.  Thus, the observed data are 
\begin{equation*}
\mathcal{D} = 
\left\lbrace \mathcal{T}_i\right\rbrace_{i=1}^n = 
\left\lbrace\left(
\bX_{i}^1, A_{i}^1, \bX_i^2, A_i^2,\ldots, 
\bX_i^K, A_i^K, \bY_i
\right)
\right\rbrace_{i=1}^n,
\end{equation*}
which comprises $n$ i.i.d. copies of the
trajectory $\mathcal{T} \triangleq (\bX^1, A^1, \bX^2, \bA^2, \ldots,
\bX^K, A^K, \bY)$, where
$\bX^1 \in\mathcal{X}^1\subseteq \mathbb{R}^{p_1}$ denotes baseline information,
$A^k\in\mathcal{A}^k$ denotes the treatment assigned
in stage $k=1,\ldots, K$, $\bX^k\in \mathcal{X}^k \subseteq \mathbb{R}^{p_k}$ denotes
information collected during the course of stage $k=2,\ldots, K$,
and $\bY\in \pmb{\mathcal{Y}} = \mathcal{Y}_1\times \mathcal{Y}_2 \times \cdot \times \mathcal{Y}_{p_y} 
\subseteq \mathbb{R}^{p_y}$ is a
vector of outcomes.   We assume that the outcomes 
$\bY= (Y_1,\ldots, Y_{p_y})$ 
have been ordered in terms of priority so that
$Y_j \in \mathcal{Y}_j$ 
is considered to be more important than $Y_{\ell}\in \mathcal{Y}_{\ell}$ when
$j < \ell$.  In our motivating study on HIV/STI prevention, 
$Y_1 \in\lbrace 0,1
\rbrace$ is an indicator of condomless anal sex with a casual 
partner (CAS),
$Y_2\in\lbrace 0 ,1\rbrace$ is an indicator of HIV testing,
and $Y_3\in \mathbb{R}$ is a measure of self-efficacy for
condom use.  We assume that all outcomes have been coded so
that higher values are preferred (e.g., in our motivating example,
we code CAS in the past three months as zero and no CAS in the past
three months as one).  

For each stage $k$, let $\bH^k \in\mathcal{H}^k$ denote the information 
available to decision makers at the time the $k$th treatment
decision is made so that $\bH^1 = \bX^1$ and
$\bH^k = (\bH^{k-1}, A^{k-1}, \bX^k)$ for $k =2,\ldots, K$. 
For each $\bh^k \in \mathcal{H}^k$ let $\psi^k(\bh^k)\subseteq
\mathcal{A}^k$ denote the (assumed to be non-empty) 
set of feasible treatments for a patient
with history $\bH^k=\bh^k$ at stage $k$ \cite[][]{van2007causal,tsiatis2019dynamic}. 
A feasible treatment regime is a sequence of functions
$\bpi = (\pi^1,\ldots, \pi^K)$ such that
$\pi^k:\mathcal{H}^k\rightarrow \mathcal{A}^k$ with 
$\pi^k(\bh^k) \in \psi^k(\bh^k)$ for all $\bh^k\in\mathcal{H}^k$.  
Let $\bPi$ denote the set of feasible treatment regimes.  For any 
$\bpi \in \bPi$ let $V_\ell(\bpi)$ denote the mean 
of outcome $Y_\ell$, $\ell = 1,\ldots, p_y$, if treatments are assigned
according to $\bpi$; let $V_\ell(\bh^1, \bpi)$ denote the mean of
outcome $Y_\ell$ conditionally on the patient presenting initially
with history $\bH^1=\bh^1$ and receiving treatment according to $\bpi$.  
A formal construction of these mean outcomes
using potential outcomes
and conditions for identifiability in terms of the data-generating model 
are provided in the Supplemental Material; throughout, we implicitly
assume these identifiability conditions hold.  We assume there exist 
clinical thresholds $\bdelta = (\delta_1,\ldots, \delta_{p_y})$ and smooth
dissimilarity 
functions $\mathbf{d} = (d_1,\ldots, d_{p_y})$ with $d_{\ell}:\mathbb{R}\times \mathbb{R}
\rightarrow \mathbb{R}_+$
such that for a patient with $\bH^1=\bh^1$ the two regimes $\bpi, \widetilde{\bpi} \in \bPi$ are clinically
indistinguishable with respect to the $\ell$th outcome if 
$d_{\ell}\left\lbrace
V_\ell(\bh^1, \bpi), V_{\ell}(\bh^1, \widetilde{\bpi})
\right\rbrace \le \delta_\ell$.  Common measures of dissimilarity include the 
absolute difference
$d_{\ell}(u, v) = |u-v|$ and the log ratio
$d_{\ell}(u, v) = \log(u/v)$ for positive values
\citep[][]{liang2020relative}.  Hereafter, we assume implicitly that
$d_\ell$ is isometric to Euclidean distance on $\mathbb{R}$  (possibly
after some transformation on the $\ell$th outcome). This condition
is not restrictive (it is clearly satisfied by the two measures noted
previously); it merely serves to avoid 
dealing with non-Euclidean geometric arguments.   

For any $\bh^1\in\mathcal{H}^1$ define the equivalence class 
of optimal regimes for outcome $\ell$ as 
\begin{equation*}
\xi_{\ell}(\bh^1) \triangleq \left\lbrace
\bpi\in\bPi\,:\, 
d_{\ell}\left[
\sup_{\widetilde{\bpi}\in\bPi} V_{\ell}(\bh^1, \widetilde{\bpi}),\,
V_{\ell}(\bh^1, \bpi)
\right] \le \delta_{\ell} 
\right\rbrace;
\end{equation*}
define the number of outcomes (ordered by priority) at which
there exists at least one optimal regime as 
\begin{equation*}
{j}(\bh^1) \triangleq \max\left\lbrace
\iota \,:\, \bigcap_{\ell=1}^\iota \xi_{\ell}(\bh^1) \ne \emptyset
\right\rbrace;
\end{equation*}
and define the set of regimes that are optimal for the 
first ${j}(\bh^1)$ outcomes as 
\begin{equation*}
\mathfrak{A}(\bh^1) \triangleq \bigcap_{\ell=1}^{{j}(\bh^1)}
\xi_{\ell}(\bh^1).  
\end{equation*}
Thus, $\mathfrak{A}(\bh^1)$ is the set of regimes one should 
consider for a patient presenting at baseline with 
$\bH^1=\bh^1$.  To make a selection within this set, we consider
two cases: (i) there exists a regime which is optimal
across all outcomes, and (ii) one must incur a clinically
meaningful loss on at least one outcome.  In case (i), one should
select a regime within the set of maximizers $\mathfrak{A}(\bh^1)$
based on the highest
priority outcome; i.e., the tie break should be based on what is
considered most important by the decision maker.  In case (ii),
one must take a clinically meaningful loss in terms of
outcome ${j}(\bh^1)+1$, thus, one should optimize
this outcome subject to being optimal on all preceding 
(higher priority) outcomes thereby minimizing negative
impact on this outcome.   
To delineate these two cases, 
define ${\tau}(\bh^1)= 1$ if 
${j}(\bh^1) = p_y$ and ${\tau}(\bh^1) = 
{j}(\bh^1) + 1$ otherwise.  Define
\begin{equation}\label{optimalRegimePreDef}
\widetilde{\bpi}(\bh^1) \triangleq \arg\max_{\bpi \in \mathfrak{A}(\bh^1)}V_{{\tau}(\bh^1)}
(\bh^1, \bpi),
\end{equation}
and define $\pmb{\pi}^{\mathrm{opt}} = (\pi^{\mathrm{opt},1},
\pi^{\mathrm{opt},2}, 
\ldots ,\pi^{\mathrm{opt},K})$ with 
$\pi^{\mathrm{opt},k}(\bh^k) 
\triangleq \left\lbrace \widetilde{\pi}^k(\bh^1)\right\rbrace\left(\bh^k\right)$, where
we have used the fact that $\bh^1$ is part of $\bh^k$ for $k\ge 1$. 
Justification for this definition is provided in the 
following two results. 

\begin{lem}
For any $\bpi\in\bPi$ and $\bh^1\in\mathcal{H}^1$, 
define the regret of
$\bpi$ on outcome $\ell$ at $\bh^1$ as   
$R_{\ell}(\bh^1, \bpi) \triangleq 
d_{\ell}\left\lbrace 
\sup_{\widetilde{\bpi}\in\bPi} V_{\ell}(\bh^1, \widetilde{\bpi}),
V_{\ell}(\bh^1, \bpi)
\right\rbrace$.  
If there exists $\bpi\in\bPi$,
$\bh^1\in\mathcal{H}^1$, and $\ell\in\left\lbrace 1,\ldots, \tau(\bh^1)
\right\rbrace$ such that  $V_{\ell}(\bh^1, \bpi) > V_{\ell}(\bh^1, 
\bpi^{\mathrm{opt}})$ and
$d_{\ell}\left\lbrace
V_\ell(\bh^1, \bpi), V_\ell(\bh^1, \bpi^{\mathrm{opt}})
\right\rbrace > \delta_{\ell}$, then $\ell > 1$ and there 
exists $\ell' < \ell$ such that 
\begin{equation*}
R_{\kappa}(\bh^1, \bpi^{\mathrm{opt}}) \le \delta_{\kappa}
\end{equation*}
for all $\kappa \le \ell'$, while   
\begin{equation*}
R_{j}(\bh^1, \bpi) > \delta_{j},  
\end{equation*}
for some $j \le \ell'$. 
That is, any regime which is clinically significantly better than
$\bpi^{\mathrm{opt}}$ on an outcome $\ell$ must incur a clinically 
significant loss on a higher priority outcome $\ell' < \ell$.  
\end{lem}

The prioritization of the outcomes can be viewed as a 
binary quasi-lexicographical ordering on the regrets
as follows.\footnote{The qualifier `quasi' here refers to the fact
that near-optimal marginal mean outcomes are considered to be equivalent
to an optimal marginal mean outcome.}  
For any regime $\bpi \in\bPi$, first stage history 
$\bh^1\in\mathcal{H}^1$, and outcome $\ell$, define  the binary 
vector $\mathbf{B}(\bh^1, \bpi) = \left\lbrace 
B_1(\bh^1, \bpi), \ldots, B_{p_y}(\bh^1, \bpi)
\right\rbrace \in \left\lbrace 0, 1\right\rbrace^{p_y}$ 
with $B_\ell(\bh^1, \bpi) = 
1_{R_{\ell}(\bh^1, \bpi) \le \delta_\ell}$.  Assume
that $\sup_{\bh_1, \bpi} R_{\ell}(\bh^1, \bpi)$ is finite
and strictly positive (see Section 4). 
For any two regimes $\bpi, \bpi'\in\bPi$, we say that $\bpi$ is preferred
to $\bpi'$ at $\bh^1$, written  
$\bpi \succeq_{\bh^1} \bpi'$, 
if either (i) there exists outcome $\kappa$ such that 
$B_{\ell}(\bh^1, \bpi) = B_{\ell}(\bh^1, \bpi') = 1$ for 
all $\ell < \kappa$, $B_{\kappa}(\bh^1, \bpi) = B_{\kappa}(\bh^1,\bpi') = 0$,
and  
$R_{\kappa}(\bh^1, \bpi) < R_{\kappa}(\bh^1, \bpi')$, 
or (ii) there exists outcome $\kappa$ such that
$B_{\ell}(\bh^1, \bpi) = B_{\ell}(\bh^1, \bpi') = 1$ for all $\ell < \kappa$ 
and $B_{\kappa}(\bh^1, \bpi) < B_{\kappa}(\bh^1, \bpi')$.   
Such 
a preference can be expressed through the utility function 
\begin{equation*}
\mathcal{U}_{\pmb{\omega}}(\bh^1, \bpi) = 
\sum_{\ell=1}^{p_y+1}\left[
\omega_\ell B_{\ell}(\bh^1, \bpi)
- \frac{
  \omega_{\ell-1}R_{\ell}(\bh^1, \bpi)
}{
  \sup_{\widetilde{\bh}^1, \widetilde{\bpi}}R_{\ell}
  (\widetilde{\bh}^1, \widetilde{\bpi})
}
\left\lbrace 
1-B_{\ell}(\bh^1, \bpi)
\right\rbrace 
\right]\prod_{\kappa=1}^{\ell-1} B_{\kappa}(\bh^1, \bpi)
,
\end{equation*} 
where $\omega_0 \equiv 1$, $\omega_{p_y+1} \equiv 1$, 
$\omega_1 > \omega_2 >\cdots > \omega_{p_y} > \omega_{p_y+1}$,
and $R_{p_y+1}(\bh^1, \bpi) \equiv R_{p_y}(\bh^1, \bpi)$; 
in addition, 
we have defined  
$\prod_{\kappa=1}^{0}B_{\kappa}(\bh^1, \bpi)
\equiv 0$; i.e., $\bpi \succeq_{\bh^1} \bpi'$ 
if and only if 
$\mathcal{U}_{\pmb{\omega}}(\bh^1, \bpi) \ge 
\mathcal{U}_{\pmb{\omega}}(\bh^1, \bpi')$.  The
following result states this more formally. 

\begin{lem}\label{preferenceLemma}
The optimal regime $\bpi^{\mathrm{opt}}$ satisfies
$\bpi^{\mathrm{opt}} \succeq_{\bh^1} \bpi$ for all
$\bh^1\in\mathcal{H}^1$ and 
$\bpi \in \bPi$. Therefore,   
\begin{equation*}
\mathcal{U}_{\pmb{\omega}}(\bh^1, \bpi^{\mathrm{opt}}) 
\ge \mathcal{U}_{\pmb{\omega}}(\bh^1, \bpi) 
\end{equation*}
for all $\bpi\in\bPi$, $\bh^1\in\mathcal{H}^1$, and all
$\pmb{\omega} \in \mathbb{R}_+^{p_y+2}$ such that
$\omega_1 > \omega_2 > \cdots > \omega_{p_y+1} = \omega_0 = 1$.  
\end{lem}

\begin{rmrk}
As noted in the introduction, the win ratio is a composite
outcome which is commonly used to compare (fixed) treatments
for a single time point under prioritized outcomes.  In a
classical application of the win ratio, 
two outcomes are typically considered to be incomparable 
if they are missing, e.g., due to censoring 
\citep[][]{finkelstein1999combining,pocock2012win}, however,
we can also say that they are not comparable if they are within
a clinically insignificant margin.  Let 
$\vartheta_{\ell}$ denote clinically margin such that if
$d_{\ell}(y_\ell, y_\ell') \le \vartheta_\ell$ their difference
is considered to be clinically insignificant.  Let 
$\bpi, \widetilde{\bpi} \in \bPi$ denote two regimes and
let $\bY^*(\bpi)$ and $\bY^*(\widetilde{\bpi})$ denote two
independent outcomes drawn from the distributions induced
by $\bpi$ and $\widetilde{\bpi}$.  One can define the 
win ratio of $\bpi$ relative to $\widetilde{\bpi}$ as follows:
\begin{multline*}
\mathrm{WR}(\bpi, \widetilde{\bpi}) = 
P\left[
Y_1(\bpi) \ge \widetilde{Y}_1(\widetilde{\bpi})
\bigg| d_{1}\left\lbrace
Y_1^*(\bpi), \widetilde{Y}_1^*(\widetilde{\bpi})
\right\rbrace > \vartheta_1
\right] \\ + 
P\left[
  Y_2^*(\bpi) \ge \widetilde{Y}_2^*(\widetilde{\bpi}) \bigg|
  d_{1}\left\lbrace
Y_1^*(\bpi), \widetilde{Y}_1^*(\widetilde{\bpi})
\right\rbrace \le \vartheta_1,\,
d_{2}\left\lbrace
Y_2^*(\bpi), \widetilde{Y}_2^*(\widetilde{\bpi})
\right\rbrace > \vartheta_2
\right] \\ 
\cdots \\ 
+ P\Bigg[
Y_{p_y}^*(\bpi) \ge \widetilde{Y}_{p_y}^*(\widetilde{\bpi}) 
\Bigg|
  d_{1}
  \left\lbrace
      Y_1^*(\bpi), \widetilde{Y}_1^*(\widetilde{\bpi})
  \right\rbrace 
       \le \vartheta_1, 
  d_{2}
  \left\lbrace 
      Y_2^*(\bpi), \widetilde{Y}_2^*(\widetilde{\bpi})
  \right\rbrace \le \vartheta_2, \\ 
\ldots,  
d_{p_y-1}\left\lbrace
Y_{p_y-1}^*(\bpi), \widetilde{Y}_{p_y-1}^*(\widetilde{\bpi})
\right\rbrace \le \vartheta_{p_y-1},
d_{\ell}\left\lbrace
Y_{p_y}^*(\bpi), \widetilde{Y}_{p_y}^*(\widetilde{\bpi})
\right\rbrace > \vartheta_{p_y}
\Bigg]. 
\end{multline*}
Thus, $\mathrm{WR}(\bpi, \widetilde{\bpi})$ captures the
probability that a randomly selected patient from
the target population recommended  
treatment under $\bpi$ will have a higher value 
than a randomly selected patient from the target 
population recommended treatment $\widetilde{\bpi}$ 
under at the first outcome at which the two are
clinically significantly different.   E.g., one
might say $\bpi$ is preferred to $\widetilde{\pi}$ if
$\mathrm{WR}(\bpi, \widetilde{\bpi}) > 1/2$. 
One could 
then attempt to build a notion of optimality for a regime
through these comparisons, however, this is
non-trivial as the win ratio need not be transitive so
there need not be a unique best regime with this definition.  
\end{rmrk}

\subsection{Inverse reinforcement learning}
\label{ref:IRLEstimand}
The regime $\bpi^{\mathrm{opt}}$ optimizes utility
as captured by the functions 
introduced in Lemma (\ref{preferenceLemma}).  
However, the relative importance of the outcomes
in $\bpi^{\mathrm{opt}}$ is not clear. Furthermore,
because the construction of $\bpi^{\mathrm{opt}}$ 
does not maximize a (known and single) 
scalar outcome measure, it is not obvious how to
conduct evaluations or sensitivity analyses; e.g.,
to compare the estimated optimal regime with
another fixed regime or to assess the importance
of a predictor on the learned decision rule
\citep[][]{sundararajan2020many}.   For the purpose
of exploratory analyses and model evaluation we
use inverse reinforcement learning \citep[][]{arora2021survey} 
to estimate a
composite outcome under which $\bpi^{\mathrm{opt}}$
is (approximately) optimal.  
Let $\varrho:\mathrm{dom}\,\mathcal{T}
\rightarrow \mathbb{R}^d$ be a fixed feature vector
defined over the space of trajectories. We
consider composite outcomes of the form
${C}_{\blambda} = 
\varrho(\mathcal{T})^{\T}\blambda$ 
indexed $\blambda \in \bLambda\subseteq\mathbb{R}^d$;
i.e., under $\blambda$, the composite
outcome (utility) for 
a patient with trajectory $\mathcal{T}=\mathbf{t}$
is  $\varrho(\mathbf{t})^{\T}\blambda$.  

For each $\blambda\in\bLambda$ and $\bpi\in\bPi$, let 
$V_{\blambda}(\bpi)$ denote the mean of 
${C}_{\blambda}$ if all patients were assigned
treatment according to $\bpi$. Let
$\bpi_{\blambda}^{\mathrm{opt}}$ be an optimal
regime for $C_{\blambda}$, i.e.,
$V_{\blambda}\left(
\bpi_{\blambda}^{\mathrm{opt}}
\right)
\ge V_{\blambda}(\bpi)
$ for all $\bpi\in\bPi$.  It can be seen that there
always exists a feature mapping $\varrho$ and
a parameter vector $\blambda$ such that 
$\bpi^{\mathrm{opt}} = \bpi_{\blambda}^{\mathrm{opt}}$.
However, in practice such a feature mapping is unknown and, 
in any case, one may wish to consider composite outcomes
that are restricted to be clinically meaningful; e.g.,
in our application we consider the feature mapping
$\varrho(\mathcal{T}) = \bY$, where each component of 
$\bY$ has been centered and
scaled.  
Assume that $\bLambda$ contains the unit ball centered at zero. 
Define 
\begin{equation*}
\blambda^{\mathrm{opt}}(\bpi^{\mathrm{opt}}) \triangleq
\arg\max_{\blambda\,:\, ||\blambda||=1} 
V_{\blambda}(\bpi^{\mathrm{opt}}),
\end{equation*}
so that $\blambda^{\mathrm{opt}}(\bpi^{\mathrm{opt}})$ indexes
the composite outcome for which $\bpi^{\mathrm{opt}}$ is
`most optimal.'   
Similar criteria have been used in 
inverse reinforcement learning 
\citep[e.g.,][]{hadfield2017inverse}.  We 
discuss estimation of $\piOpt$ and
$\lamOpt(\piOpt)$ in the next section.  


\section{Estimation}\label{sec:Estimation}
\subsection{Estimation of $\piOpt$}
Our estimator uses regression-based
policy-search to construct an
estimator of $\piOpt$  
\citep[][]{listBasedRegimes,tsiatis2019dynamic}.  
Let $\bpi\in\Pi$ be arbitrary and for any
$k$ write $\underline{\bpi}^k \triangleq
(\pi^k, \pi^{k+1},\ldots, \pi^K)$.  
Define
$\bQ^K(\bh^K, a^K) \triangleq 
\mathbb{E}\left(
\bY\big|\bH^K=\bh^K, A^K=a^K
\right)$ and 
\begin{equation*}
\bQ^{K-1}(\bh^{K-1}, a^{K-1}; \pi^K)
= \mathbb{E}\left[
\bQ^K\left\lbrace
\bH^K, \pi^K(\bH^K)
\right\rbrace \bigg| \bH^{K-1}=\bh^{K-1},
A^{K-1}=a^{K-1}
\right].
\end{equation*}
Thus, $\bQ^K(\bh^K, a^K)$ is the mean vector of 
outcomes for
a patient presenting at the last stage with
$\bH^K=\bh^K$ and assigned to treatment
$A^K=a^K$, while $\bQ^{K-1}(\bh^{K-1}, a^{K-1};
\pi^{K})$ is the mean outcome for a patient
presenting at the penultimate stage with
$\bH^{K-1}=\bh^{K-1}$, assigned treatment
$A^{K-1}=a^{K-1}$ and then treated according
to $\pi^K$ at the final stage.  
Recursively, for $k=K-2,\ldots, 1$ define
\begin{equation*}
\bQ^k(\bh^k, a^k; \underline{\bpi}^{k+1}) \triangleq
\mathbb{E}\left[
\bQ^{k+1}\left\lbrace 
\bH^{k+1}, \pi^{k+1}(\bH^{k+1});
\underline{\bpi}^{k+2}
\right\rbrace
\big|\bH^k=\bh^k, A^k=a^k
\right].
\end{equation*}
It can be seen that $\bQ^k(\bh^k, a^k; \underline{\bpi}^{k+1})$ is 
the mean outcome for a patient presenting with
$\bH^k=\bh^k$, assigned treatment $A^k=a^k$, and then
assigned treatment according to $\underline{\bpi}^{k+1}$
thereafter.  Therefore, for any
$\ell =1,\ldots, p_y$, it follows that
$V_{\ell}(\bh^1, \bpi) = Q_\ell^1
\left\lbrace \bh^1, \pi^1(\bh^1);\underline{\bpi}^2
\right\rbrace$, so that
\begin{equation*}
\xi_{\ell}(\bh^1) = \left\lbrace
\bpi\in\bPi \,:\, 
d_{\ell}\left[
\sup_{\widetilde{\bpi}\in\bPi} 
Q_{\ell}^1\left\lbrace
\bh^1, \widetilde{\pi}^{1}(\bh^1); 
\underline{\widetilde{\bpi}}^{2}
\right\rbrace,\,
Q_{\ell}^1\left\lbrace
\bh^1, \pi^1(\bh^1);
\underline{\bpi}^2
\right\rbrace
\right] \le \delta_{\ell}
\right\rbrace,
\end{equation*}
which shows that the first stage
$Q$-function, $\bQ^1$, 
determines $\widetilde{j}(\bh^1)$,
$\mathfrak{A}(\bh^1)$, and thus
the optimal regime 
$\piOpt$  (per Equation (\ref{optimalRegimePreDef})).

Let $\mathcal{Q}^K$ denote a generic model class 
for $\bQ^K(\bh^K, a^K) = \mathbb{E}(\bY|\bH^K=\bh^K,
A^K=a^K)$, e.g., one might use a low-dimensional 
parametric model, or a more flexible machine
learning method such as  gradient boosting or 
neural networks, etc.
Let $\widehat{\bQ}_n^K(\bh^K, a^K) \in \mathcal{Q}^K$
denote an estimator of $\bQ^K$ constructed from
$\mathcal{D}$.  For example, $\widehat{\bQ}_n^K(\bh^K, a^K)$ might solve 
\begin{equation*}
\widehat{\bQ}_n^K = \arg\min_{\mathbf{Q}\in\mathcal{Q}^K}
\pn ||\bY - \bQ(\bH^K, A^K)||^2,
\end{equation*}
where $\pn$ denotes the empirical measure.  
For $k=K-1,K-2,\ldots, 1$ 
let $\widehat{\bQ}_n^k(\bh^k, a^k; \underline{\bpi}^{k+1})\in\mathcal{Q}^k$ be an estimator
of $\bQ^k(\bh^k, a^k; \underline{\bpi}^{k+1})$, 
where $\mathcal{Q}^k$ is a posited class of
models.  For example, $\widehat{\bQ}_n^k(\bh^k, a^k; \underline{\bpi}^{k+1})$ 
might solve
\begin{equation*}
\widehat{\bQ}_n^k = \arg\min_{\bQ\in\mathcal{Q}^k}
\pn ||\widehat{\bQ}_n^{k+1}\left\lbrace 
\bH^{k+1}, {\pi}^{k+1}(\bh^{k+1}); \underline{\bpi}^{k+2}
\right\rbrace
- \bQ(\bH^k, A^k; \underline{\bpi}^{k+1})
||^2,
\end{equation*}
where, for notational consistency, 
we have defined $\widehat{\bQ}_n^K(\bh^K, a^K; \underline{\bpi}^{K+1}) 
\equiv \widehat{\bQ}_n^K(\bh^K, a^K)$.  

The plug-in estimator of $V_{\ell}(\bh^1, \bpi)$ is
$\widehat{V}_{\ell,n}^1(\bh^1,  \bpi) = 
\widehat{Q}_{\ell,n}^1\left\lbrace
\bh^1, \pi^1(\bh^1); \underline{\bpi}^2
\right\rbrace$. Subsequently, the estimated equivalence
class of optimal regimes for outcome $\ell$ is 
\begin{equation*}
\widehat{\xi}_{\ell,n}(\bh^1) \triangleq \left\lbrace
\bpi\in\bPi\,:\, d_{\ell}\left[
\sup_{\widetilde{\bpi}\in\bPi} \widehat{V}_{\ell,n}(\bh^1, \widetilde{\bpi}),\,
\widehat{V}_{\ell,n}(\bh^1, \bpi)
\right] \le \delta_{\ell}
\right\rbrace,
\end{equation*}
and the estimated number of outcomes (ordered by priority) for which there
exists at least one optimal regime is
\begin{equation*}
\widehat{{j}}_n(\bh^1) = \max\left\lbrace
\iota \,:\, \bigcap_{\ell=1}^{\iota}\widehat{\xi}_{\ell,n}(\bh^1) \ne \emptyset 
\right\rbrace;
\end{equation*}
and the estimated optimal set of regimes is 
\begin{equation*}
\widehat{\mathfrak{A}}_n(\bh^1) = \bigcap_{\ell=1}^{\widehat{{j}}_n(\bh^1)}
\widehat{\xi}_{\ell,n}(\bh^1).  
\end{equation*}
Letting $\widehat{{\tau}}_n(\bh^1) = 1$ if 
$\widehat{{j}}_n(\bh^1) = p_y$ and 
$\widehat{{\tau}}_n(\bh^1) = \widehat{{j}}_n(\bh^1) + 1$
otherwise, and 
$\widehat{\widetilde{\bpi}}(\bh^1) = \arg\max_{\bpi\in\widehat{\mathfrak{A}}_n(\bh^1)}
\widehat{V}_{\widehat{{\tau}}_n(\bh^1) ,n}(\bh^1,\bpi)$, the estimated 
optimal regime is 
$\widehat{\pi}_{n}^k(\bh^k) = \left\lbrace
\widehat{\widetilde{\bpi}}_n(\bh^1)
\right\rbrace^k (\bh^k).$

%

\subsection{Estimation of $\lamOpt(\piOpt)$} \label{sec:est_lam}
We use a regression-based estimator of 
$\lamOpt(\piOpt)$ based on the characterization in 
Section \ref{ref:IRLEstimand}.  
Let $\mathcal{Q}_{\blambda}^K$ denote a class of 
models for $Q_{\blambda}^K(\bh^K, a^K; \piOpt) = 
\mathbb{E}\left(
C_{\blambda}\big|\bH^{K}=\bh^K, A^K=a^K
\right)$ and let $\widehat{Q}_{\blambda,n}^K(\bh^K, a^K)$ denote
an estimator of $Q_{\blambda}^K(\bh^K, A^K;\piOpt)$ constructed
via least squares 
\begin{equation*}
\widehat{Q}_{\blambda,n}^{K} = 
\arg\min_{Q\in\mathcal{Q}_{\blambda}^K}
\pn \left\lbrace
C_{\blambda} - Q(\bH^K, A^K)
\right\rbrace^2.
\end{equation*}
Similarly, let $\mathcal{Q}_{\blambda}^k$ 
denote a class of
models for 
\begin{equation*}
Q_{\blambda}^k(\bh^k, a^k;\piOpt) = 
\mathbb{E}\left[
Q_{\blambda}^{k+1}\left\lbrace
\bH^{k+1}, \pi^{\mathrm{opt},k+1}(\bH^{k+1});
\piOpt
\right\rbrace\big|\bH^{k}=\bh^{k},
A^{k}=a^{k}
\right]
\end{equation*}
and 
define 
\begin{equation*}
\widehat{Q}_{\blambda,n}^{k} = 
\arg\min_{Q\in\mathcal{Q}_{\blambda}^{k}}
\pn \left[
\widehat{Q}_{\blambda,n}^{k+1}\left\lbrace
\bH^{k+1}, \widehat{\pi}_{n}^{k+1}(\bH^{k+1})
\right\rbrace
-Q(\bH^k, A^k)
\right]^2,
\end{equation*}
so that $\widehat{Q}_{\blambda,n}^k(\bh^k, a^k)$
is the resulting estimator of
$Q_{\blambda}^{k}(\bh^k, a^k;\piOpt)$.  The plug-in
estimator of $V_{\blambda}(\bpi^{\mathrm{opt}})$ is
$\widehat{V}_{\blambda,n} = \pn 
\widehat{Q}_{\blambda,n}^{1}\left\lbrace
\bH^1, \widehat{\pi}_{n}^1(\bH^1)
\right\rbrace$, and subsequently the estimator
of $\lamOpt(\piOpt)$ is 
$\widehat{\blambda}_{n} = 
\arg\max_{\blambda\,:\,||\blambda||=1}\widehat{V}_{\blambda,n}$.

\section{Inference for the value of an estimated regime}\label{sec:Theory}
In our motivating application of adaptive educational interventions
targeting HIV/STI prevention, as well as many other precision public
health applications, 
the performance of an estimated optimal treatment regime is of primary interest.  
In this
section, we describe how to use sample splitting to estimate
an optimal treatment regime and to evaluate its 
performance.   Unless
otherwise stated, the proposed methods do not impose any conditions on the
estimators of the Q-functions save for consistency. 

To simplify notation, we assume that the number of subjects 
is even, $n = 2m$, and that the data have been randomly partitioned into 
$\mathcal{D}^{(1)} = 
\left\lbrace (\bX_i^1, A_i^1, \bX_i^2,A_i^2,\ldots,
\bX_i^K, A_i^K, \bY_i^K)\right\rbrace_{i=1}^m$ 
and $\mathcal{D}^{(2)} = \left\lbrace (\bX_i^1, A_i^1, \bX_i^2,A_i^2,\ldots,
\bX_i^K, A_i^K, \bY_i^K)\right\rbrace_{i=m+1}^{2m}$.  
Let $\mathbb{P}_m^{(j)}$ denote the empirical
measure on $\mathcal{D}^{(j)}$ and for any statistic
$\widehat{T}_n = T(\pn)$ write
$\widehat{T}_m^{(j)} = T(\mathbb{P}_m^{(j)})$ to 
denote the statistic computed using $\mathcal{D}^{(j)}$.  
Thus, $\widehat{\bpi}_m^{(j)}$ is the estimated
optimal strategy using $\mathcal{D}^{(j)}$.  In what 
follows, we estimate  the optimal treatment
regime using $\mathcal{D}^{(1)}$ and evaluate
it using the augmented inverse probability weighted
estimator \citep[AIPWE][]{tsiatis2019dynamic} 
computed on $\mathcal{D}^{(2)}$.  
For any regime $\bpi$ and $k=1,\ldots, K$ define coarsening variable
\begin{equation*}
\zeta_{\bpi}^k = 
\prod_{\ell=1}^{k}1_{A^{\ell} = \bpi^\ell(\bH^\ell)},
\end{equation*}
so that $\zeta_{\bpi}^k$ is an indicator that 
$(\bH^k, A^k)$ is consistent with $\bpi$ 
for the first $k$ stages.  
Define the sample-splitting AIPWE of 
$\bV(\widehat{\bpi}_m^{(1)})$
as 
\begin{multline*}
\widehat{\bV}_m^{(2)}(\widehat{\bpi}_m^{(1)}) = 
\mathbb{P}_m^{(2)}\left\lbrace
\frac{
  \bY\zeta_{\widehat{\bpi}_{m}^{(1)}}^K
}{
  \prod_{\ell=1}^KP(A^\ell|\bH^\ell)
}\right\rbrace 
\\ + \,\, \mathbb{P}_{m}^{(2)}\sum_{k=1}^K
\left\lbrace
\frac{
  \zeta_{\widehat{\bpi}_{m}^{(1)}}^{k-1}
}{
  \prod_{\ell=1}^{k-1}P(A^\ell|\bH^\ell)
} -
\frac{
  \zeta_{\widehat{\bpi}_{m}^{(1)}}^{k}
}{
  \prod_{\ell=1}^{k}P(A^\ell|\bH^\ell)
} 
\right\rbrace 
\widehat{\bQ}_{m}^{k(1)}\left\lbrace
\bH^{k}, \widehat{\bpi}_{m}^{(1)}(\bH^k);
\underline{\widehat{\bpi}}_{m}^{k+1(1)}
\right\rbrace.  
\end{multline*}
Our goal is to construct valid conditional 
confidence regions 
for $\bV(\widehat{\bpi}_m^{(1)})$ given
$\mathbb{P}_m^{(1)}$.  
We will make use of
the following assumptions.  
\begin{itemize}
\item[(A0)] {\em Standard causal assumptions.}  Positivity, 
strong ignorability, and consistency hold. See the Supplemental
Materials for a formal statement of these assumptions. 
\item[(A1)] {\em H\"older distance and convergence on compacta.}  
For each $\ell=1,\ldots, L$, $d_{\ell}$ is 
H\"older, i.e., $d_{\ell}$ is 
a metric and there exists $\varsigma_{\ell}, C_{\ell} > 0$ such that 
$d_{\ell}(u,v) \le C_{\ell}|u-v|^{\varsigma_{\ell}}$ for all $u,v\in\mathbb{R}$.  
For any compact set
$\mathcal{C}\subseteq \mathcal{H}^1$ 
\begin{equation*}
\sup_{\bpi\in\bPi}\big|
\widehat{V}_{\ell,n}(\bh^1, \bpi) - V_{\ell}(\bh^1, \bpi)
\big| \rightarrow_{p} 0,
\end{equation*}
for each $\bh^1 \in \mathcal{C}$ and $\ell=1,\ldots, L$.    
\item[(A2)] {\em Consistency and equicontinuity 
of the estimated Q-functions.}  
For 
each $k=1,\ldots, K$ the
estimated $Q$-functions satisfy 
\begin{equation*} 
\sup_{\bpi\in\Pi} \sup_{\bh^k\in\mathcal{H}^k} \big|\big|
\widehat{\bQ}_n^k\left\lbrace 
\bh^k, \pi^k(\bh^k); \underline{\bpi}^{k+1})\right\rbrace 
- 
\bQ^k\left\lbrace \bh^k, \pi^k(\bh^k); \underline{\bpi}^{k+1}\right\rbrace
\big|\big|
\rightarrow_p 0,
\end{equation*}
as $n\rightarrow\infty$. 
For any $\bh^1\in\mathcal{H}^1$ and $\bpi,\bpi\in\bPi$, define
\begin{equation*}
\mathfrak{d}(\bpi, \bpi'; \bh^1) = 
\mathbb{E}\left\lbrace \Bigg|\Bigg|
\frac{
    \bY\zeta_{\bpi}^K
}{
  \prod_{j=1}^K P(A^j|\bH^j)
}
- 
\frac{\bY\zeta_{\bpi'}^K
}{
  \prod_{j=1}^K P(A^j|\bH^j)
}
\Bigg|\Bigg|^2 \,\Bigg| \bH^1=\bh^1
\right\rbrace.  
\end{equation*}
Given 
$\varepsilon > 0$ there exists $\mathfrak{b} > 0$ such that 
\begin{equation*}
\sup_{\bpi\in\bPi \atop \mathfrak{d}\left(
\bpi, \bpi^{\mathrm{opt}}; \bh^1
\right) > \varepsilon} 
Q_{\tau(\bh^1)}^1\left(\bh^1, \bpi\right) \le 
Q_{\tau(\bh^1)}^1\left(
  \bh^1, \bpi^{\mathrm{opt}}
\right) - \mathfrak{b}.  
\end{equation*}

\item[(A3)] {\em Positive definite asymptotic covariance.} 
Define 
\begin{equation*}
\bSigma \triangleq P\left\lbrace
\frac{
  \bY\zeta_{\bpi^{\mathrm{opt}}}^K
}{
  \prod_{j=1}^K P(A^j|\bH^j)
} - P \frac{
  \bY\zeta_{\bpi^{\mathrm{opt}}}^K
}{
  \prod_{j=1}^K P(A^j|\bH^j)
} 
\right\rbrace
\left\lbrace
\frac{
  \bY\zeta_{\bpi^{\mathrm{opt}}}^K
}{
  \prod_{j=1}^K P(A^j|\bH^j)
} - P \frac{
  \bY\zeta_{\bpi^{\mathrm{opt}}}^K
}{
  \prod_{j=1}^K P(A^j|\bH^j)
} 
\right\rbrace^{\T},
\end{equation*}
then 
$\bSigma$ is positive definite.  
\item[(A4)] {\em Bounded utility.}  We assume that $\mathcal{Y}$ is a bounded
subset of $\mathbb{R}^{p_y}$.  
\item[(A5)] {\em Margin conditions.}  For each $\ell=1,\ldots, p_y$ 
\begin{equation*}
\sup_{\bpi\in\bPi} 
P\left(\bigg|
d_{\ell}\left[
\sup_{\widetilde{\bpi}\in\bPi}V_{\ell}(\bH^1, \widetilde{\bpi}),
V_{\ell}(\bH^1, \bpi)
\right] - \delta_{\ell}\bigg| \le  \eta
\right) \rightarrow 0,
\end{equation*}
as $\eta \rightarrow 0$.  
\end{itemize}
\noindent 
These assumptions are quite mild.  (A0) is assumed as a matter of course of sequential
decision problems \citep[][]{tsiatis2019dynamic}; (A1) imposes no conditions on the
rate of convergence and thus allows for the use of flexible estimators, e.g.,
neural networks, boosting, random forests, etc.; (A2) similarly 
assumes that the estimated $Q$-functions are consistent
and, as is standard in $M$-estimation, that the there is a unique isolated optimum;  (A3) is a standard regularity condition; 
(A4) holds in health settings where $\bY$ comprises physical or
behavioral measurements like those in the HIV/STI prevention study which
motivates this work; and (A5) ensures $\bpi^{\mathrm{opt}}$ is uniquely
identified 
\cite[e.g., see][]{luedtke2016statistical,rose2019sample,shi2020statistical,ertefaie2021robust}. The following result is proved
in the Supplemental Materials. 
\begin{thm}
Assume (A0)-(A4) and let $\Phi$ denote the cumulative distribution 
function of a standard normal random variable.  For any 
$\bc\in\mathbb{R}^{p_y}$ such that $||\bc|| = 1/\sqrt{p_y}$, 
it follows
that
\begin{equation*}
\sup_{t\in\mathbb{R}}\bigg|
P\left[
c^{\T}\bSigma^{-1/2}\sqrt{m}\left\lbrace
\widehat{\bV}_{m}^{(2)}(\widehat{\bpi}_m^{(1)}) - 
\bV(\widehat{\bpi}_m^{(1)})
\right\rbrace  \le t 
\right] - \Phi(t)
\bigg| \rightarrow_p 0,
\end{equation*}
as $m\rightarrow \infty$.
\end{thm}
\noindent 
The preceding result establishes consistency and asymptotic 
normality of the estimator 
$\widehat{\bV}_{m}^{(2)}(\widehat{\bpi}_{m}^{(1)})$.  However, to construct
confidence sets for $\bV(\widehat{\bpi}_{m}^{(1)})$ requires a consistent 
estimator $\widehat{\bSigma}_m$ of $\bSigma$; hence, the following
the result.  

\begin{lem}
Assume (A0)-(A4) and define 
\begin{equation*}
\widehat{\bSigma}_m \triangleq 
\mathbb{P}_m^{(2)}
\left\lbrace
\frac{
  \bY\zeta_{\widehat{\bpi}_m^{(1)}}^K
}{
  \prod_{j=1}^K P(A^j|\bH^j)
} - \mathbb{P}_m^{(2)} \frac{
  \bY\zeta_{\widehat{\bpi}_m^{(1)}}^K
}{
  \prod_{j=1}^K P(A^j|\bH^j)
} 
\right\rbrace
\left\lbrace
\frac{
  \bY\zeta_{\widehat{\bpi}_m^{(1)}}^K
}{
  \prod_{j=1}^K P(A^j|\bH^j)
} - \mathbb{P}_m^{(2)} \frac{
  \bY\zeta_{\widehat{\bpi}_m^{(1)}}^K
}{
  \prod_{j=1}^K P(A^j|\bH^j)
} 
\right\rbrace^{\T}.
\end{equation*}   
Let $||\cdot||_F$ denote
the Frobenius norm, then for any $\eta > 0$ 
\begin{equation*}
P\left(
||\widehat{\bSigma}_{m} - \bSigma||_F > \eta 
\right) \rightarrow 0,
\end{equation*}
an $m\rightarrow\infty$.  
\end{lem}\noindent
The following is an immediate consequence
of the preceding two results.  
\begin{cor}
Assume (A0)-(A4) and let $\widehat{\bSigma}_{n}$ 
be the plug-in estimator
of $\bSigma$.  For any $\alpha \in (0,1)$, define 
\begin{equation*}
\mathfrak{M}_{m,1-\alpha} \triangleq \left\lbrace
\nu \,:\, m\left[
\widehat{\bV}_{m}^{(2)}(\widehat{\bpi}_{m}^{(1)}) - \nu 
\right]^{\T} \widehat{\bSigma}_{m}^{-1}
\left[
\widehat{\bV}_{m}^{(2)}(\widehat{\bpi}_{m}^{(1)}) - \nu 
\right] \le \chi_{p_y, 1-\alpha}^2
\right\rbrace, 
\end{equation*}
where $\chi_{p_y, 1-\alpha}^2$ is the upper $(1-\alpha)$ 
quantile of a $\chi^2$-distribution with $p_y$ degrees
of freedom.  Then $P\left\lbrace
\bV(\widehat{\bpi}_{m}^{(1)}) \in \mathfrak{M}_{m,1-\alpha}
\right\rbrace \ge 1-\alpha + o_P(1).$
\end{cor}

In addition to inference for the performance
of the estimated optimal treatment regime, one
may be interested in conducting inference for the
composite outcome weights
$\lamOpt(\piOpt)$.  The following result
constructs as asymptotic confidence set for
$\lamOpt(\piOpt)$ using a simple extension 
of universal inference 
\citep[][]{wasserman2020universal}.  
\begin{lem}
Assume (A0)-(A3).
Let $\widehat{\blambda}_{m}^{(1)}$ denote the
estimator of $\lamOpt(\piOpt)$ constructed on
$\mathcal{D}^{(1)}$. For any fixed
$\alpha \in (0,1)$ define
\begin{equation*}
\zeta_{m,\alpha} \triangleq\left\lbrace
\blambda\,:\, ||\blambda|| = 1
\mbox{ and }
\frac{
  \widehat{V}_{\blambda,m}^{(2)}
}{
  \widehat{V}_{\widehat{\blambda}_{m}^{(1)},m}^{(2)}
}
\le \frac{1}{\alpha}
\right\rbrace.
\end{equation*} 
Then $P\left\lbrace
\lamOpt(\piOpt) \in \zeta_{m,\alpha}
\big|\mathbb{P}_m^{(1)}
\right\rbrace \ge 1-\alpha + o_P(1)$.
\end{lem} \noindent

\section{Simulation experiments}\label{sec:Sims}
We use a series of simulation experiments to 
evaluate the finite-sample performance of the proposed
prioritized regime estimator along with the
adaptive composite estimator discussed
in Section \ref{sec:est_lam}. Performance
is measured in terms of the marginal mean
outcome of each clinical endpoint.  For the 
purpose of comparison, we also implemented
$Q$-learning with random forests applied separately to each
outcome as well as to a composite outcome formed by 
taking the average of all outcomes, i.e.,
$Y_c = p_y^{-1}\sum_{\ell=1}^{p_y}Y_\ell$.  
We also evaluated the coverage and width 
of the sample-splitting confidence interval developed in
the previous section.

\par We consider two classes of generative models 
based on modifying those used in \citet{laber2014statistical} and \citet{yingqi2}
to have three outcomes of interest $(Y_1, Y_2, Y_3)$ so that $p_y=3$.  
Each design has two stages of randomization with binary treatments at each
stage.  Treatments are assigned uniformly at random, 
so that $P(A^k=1)=P(A^k=-1)=1/2$ for $k=1,2$.  

The first collection of generative models is as follows. 
Let $\bepsilon\sim N(0, I_3)$ and 
$\pmb{\upsilon} \sim N(0, I_3)$. 
We generate baseline covariates $\bX^1\sim\mathrm{N}(0,I_3)$ and
 second stage covariates
 $X_j^2 = 0.5X^1_{j}A_1 + \upsilon_{j}$ for $j=1,2,3$. In addition,
 define 
\begin{equation}\label{zFormula}
   Z_j =\gamma_1 + \gamma_2X_{j}^1+\gamma_3A^1+\gamma_4X_{j}^1A_1+\gamma_5A^2+
   \gamma_6X_{j}^2A^2+\gamma_7A^1A^2+\epsilon_{j};
\end{equation}
for $k\in\{1,2,3\}$;
the values of $\gamma$ corresponding to each $Z_j$ are given in Table \ref{tab:value_params}. 
In the first collection of generative models, we consider
two outcome distributions for $Y_\ell$ for $\ell\in\{1,2,3\}$:

\begin{itemize}
    \item[(S1)] $\mathbf{Y}=\left(Z_1, Z_2, 3Z_1/3+Z_2/3+Z_3/3 \right)$ 
    with $\bdelta=(0.1, 0.1, 0.1)$;\medskip
    \item[(S2)] $\mathbf{Y}=\left(4Z_1/4+Z_2/10+Z_3/10, Z_2/5+4Z_3/4, 
    Z_1/2 + Z_2/2 + Z_3/3\right)$ with \\ $\bdelta=(0.25, 0.25, 0.25)$.
\end{itemize}

\begin{table}[h]
\caption{\label{tab:value_params}Parameters indexing the six generative models in 
(\ref{zFormula}).}
\begin{center}
\begin{tabular}{c|c}
  \hline
 Setting & $\gamma$ \\
  \hline
\text{1} & $(0,0.5, -0.7, 0.75, 0.3, 0.5, 0.5)$\\
\text{2} & $(0,0.5, 0.7, 0.75, 0.3, 0.5, 0.5)$\\
\text{3} & $(0 , 0.5, 0, 0.75, 0.1, 0.5, 0)$\\
   \hline
\end{tabular}
\end{center}
\end{table}

The second collection of generative models that we consider is as follows. 
Baseline covariates are generated as $\bX^1 \sim N(0, I_4)$
and second stage covariates are given by  
\begin{equation*}
  \begin{array}{ll}
  X^2_{1} = 1\left\{1.25 X^1_1A^1 + \upsilon_1 > 0\right\}, & 
     X^2_{2} = 1\left\{-1.75 X^1_1A^1 + \upsilon_2 > 0\right\}, \medskip \\ 
  X^2_{3} =  1 + 1.5X^1_3A^1 + \upsilon_3, &  \medskip  
    X^2_{4} = 0.5X^1_3A^1 + \upsilon_4,
  \end{array}
\end{equation*}
where $\pmb{\upsilon} \sim N(0, I_4).$

Let $\bepsilon \sim N(0, I_3)$ and define  
\begin{eqnarray*}
  Z_4 &=& 1.5A^2(0.5 + X^2_3 + 0.5A^1 + 0.5X^2_1 - 0.5X^2_2) + \epsilon_{1}, \medskip \\ 
  Z_5 &=& -1.5A^2(0.5 + X^2_3 + 0.5A^1 + 0.5X^2_1 - 0.5X^2_2) + \epsilon_{2}, \medskip \\ 
  Z_6 &=& 2A^2(0.75 - X^2_3 + 0.75A^1 - 0.75X^2_1 - 0.25X^2_2) + \epsilon_{3}.
\end{eqnarray*}

We consider two outcome settings, given by 
\begin{itemize}
    \item[(S3)] $\mathbf{Y}=\left(Z_4, Z_6, Z_5\right)$ and $\bdelta=(0.5, 0.5, 0.5)$;\medskip
    \item[(S4)] $\mathbf{Y}=\left(Z_5, Z_6, Z_4\right)$ and $\bdelta=(0.5, 0.5, 0.5)$.
\end{itemize}

Our class of decision rules at the second stage is as follows. Define
\begin{equation*}
\bPi^2 = \left\lbrace \pi^2(\bh^2;\bLambda) 
= \arg\max_{a^2\in\psi^2(\bh^2)}\sum_{\ell=1}^L\lambda_\ell 
Q_\ell^2(\bh^2, a^2)
\,:\, \blambda \in [0,1]^L \mbox{ and } \sum_{\ell=1}^L\lambda_\ell = 1
\right\rbrace,
\end{equation*}
so that $\bPi^2$ is the set of optimal decision rules for convex combinations 
of the outcome at the second stage.  Of course, $\bPi^2$ is unknown,  so we approximate it with
$\widehat{\bPi}_m^2$ which is obtained by replacing the $Q$-functions with
their estimated counterparts and by taking a large sample of $\blambda$ values
from the $p_y$-dim probability simplex (we used 1000 samples).  The
set of first stage decision rules is unrestricted save for the requirement 
that $\pi^1(\bh^1) \in \psi^1(\bh^1)$. In order to mirror the HIV/STI prevention study 
(which has $n=1024$ participants),
all simulations are conducted with sample size $n=1000$ and 
a sample split size $m=n/2=500$. Reported
results are based on $1000$ Monte Carlo
replications. The conditional value of the estimated regime is approximated using a test
set of size $10000.$

\par Table \ref{simResults} displays the results.  We see that the prioritized regime aligns 
with the stated priority and maximizes the mean utility. The composite outcome
derived from the prioritized outcome has similar performance which suggests that the 
learned weights (we consider composite outcomes of the form $\bY^{\T}\pmb{\lambda}$) 
may contain useful information about  how the outcomes are weighted by the prioritized
regime.  Q-learning (QL) for a single outcome, as expected, optimizes the targeted outcome
at the cost of poor performance across the other outcomes.  Optimizing for the average
outcome can lead to poor performance in terms of the highest priority outcome.  
The table also reports the average coverage of a 95\% confidence interval for the
value of the estimated regime; nominal coverage is attained in all settings.

\begin{table}[h!]
 \caption{\label{simResults} Average conditional value of estimated regimes across $1,000$ Monte Carlo replications and coverage probability (CP) of the sample-splitting confidence 
 interval. 
}
\begin{tabular}{l|rr|rr|rr}
Regime & $V_1$ & CP & $V_2$ & CP & $V_3$ & 
ECP  \\ \hline
Prioritized S1& $1.296$ & $0.956$ & $0.189$ & $0.943$ & $0.517$ & $0.937$\\[3pt]
$Q$-learning S1& $1.396$ & $0.936$ & $-0.063$ & $0.944$ & $0.446$ & $0.951$\\[3pt]
Composite S1& $0.791$ & $0.947$ & $1.293$ & $0.946$ & $0.743$ & $0.941$\\[3pt]
Tuned Composite S1 & $1.369$ & $0.932$ & $0.111$ &$0.936$ & $0.511$ & $0.947$
\\[3pt]
\\
Prioritized S2& $1.076$ & $0.933$ & $0.226$ & $0.937$ & $0.803$ & $0.943$\\[3pt]
Q-learning S2& $1.123$ & $0.952$ & $-0.091$ & $0.932$ & $0.772$ & $0.950$\\[3pt]
Composite S2& $0.850$ & $0.945$ & $0.665$ & $0.946$ & $1.014$ & $0.936$\\[3pt]
Tuned Composite S2 & $1.078$ &  $0.952$ & $0.255$ & $0.944$ & $0.934$ & $0.948$\\[3pt]
\\
Prioritized S3 & $2.930$ & $0.934$ & $-1.118$ & $0.940$ & $-2.930$ & $0.940$\\[3pt]
$Q$-learning S3 & $3.341$ & $0.938$ & $-2.017$ & $0.941$ &
$-3.341$ & $0.941$ \\[3pt]
Tuned Composite S3 & $3.371$ & $0.958$ & $-1.836$ &$0.948$ & $-3.371$ & $0.929$
\\[3pt]
Composite S3& $-2.156$ & $0.929$ & $3.078$ &$0.945$ & $2.153$ & $0.942$
\\[3pt]
\\
Prioritized S4 & $2.890$ & $0.953$ & $-1.378$ & $0.955$ & $1.378$ & $0.958$\\[3pt]
$Q$-learning S4 & $3.087$ & $0.932$ & $-2.097$ & $0.939$ & $2.097$ & $0.948$\\[3pt]
Tuned Composite S4 & $2.772$ & $0.933$ & $-2.547$ &$0.932$ & $2.547$ & $0.945$
\\[3pt]
Composite S4& $2.153$ & $0.942$ & $3.078$ &$0.945$ & $-2.156$ & $0.929$
\\[3pt]
\end{tabular}
\end{table}


\section{Case study}\label{sec:CaseStudy}
\begin{figure}
\begin{center}
\includegraphics[width=4in]{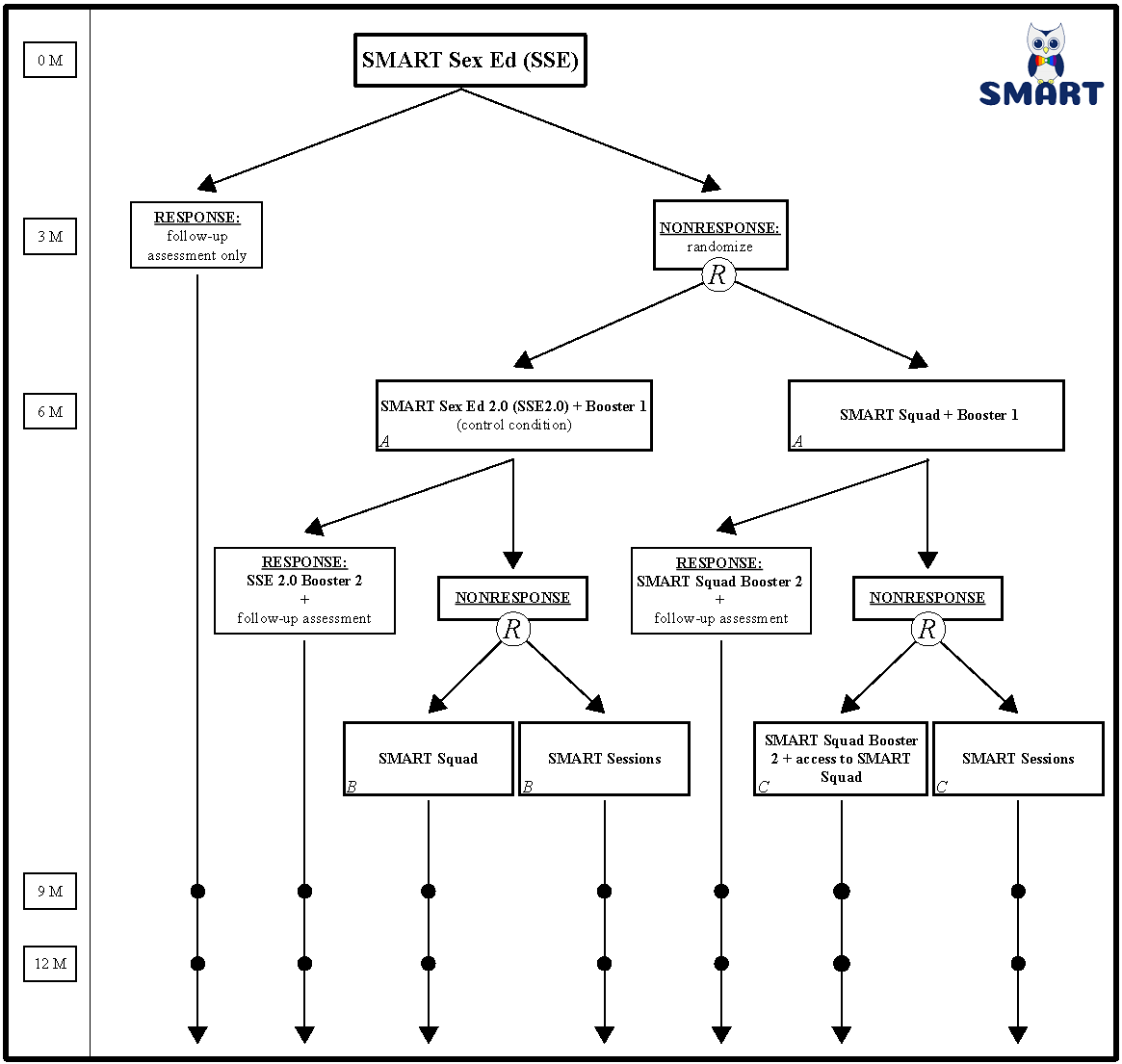}
\caption{
  \label{smartDiagram} Schematic for SMART trial evaluating
  eHealth strategies for HIV/STI prevention among
  adolescent men who have sex with men.  During the initial
  phase all participants are assigned to SMART Sex Ed (SSE).
  Non-responders to SSE are randomized with equal  probability
  to SMART Sex 2.0 with a SSE booster (SSE 2.0+) or SMART Squad
  with SSE booster (SSBooster).  In the second stage of
  randomized treatment assignment, non-responders 
  to SSE 2.0+ are randomized with equal probability
  to SMART Squad (SS) or SMART Sessions (SSess), while
  non-responders to to SS+ are randomized with equal probability to 
  SS Booster and SMART Squad (SSB) or SESS.  Responders are 
  not re-randomized.   
}
\end{center}
\end{figure}

We applied the proposed estimator to data from a 
SMART on HIV/STI prevention through eHealth in adolescent men
who have sex with men \citep[][]{mustanski2020evaluation}.  
A schematic of the trial's protocol is shown
in Figure \ref{smartDiagram}.  
The SMART contains a burn-in treatment
(SMART Sex Ed (SSE)) followed by up to two-stages
of randomization for non-responders.  In the first
stage of randomized treatment assignment, non-responders to SSE are randomized
with equal probability to one of two eHealth
interventions: SMART Sex Ed 2.0 with SSE booster 
(SSE 2.0+) or SMART Squad with SSE booster (SS+).  
In the second stage of randomized treatment assignment, 
non-responders to SSE 2.0+ are randomized with equal
probability to 
SMART Squad (SS) or SMART Sessions (SSess), while
non-responders to SS+ are randomized with equal probability
to SMART Squad Booster and SMART Squad (SSBooster) or SMART 
Sessions (SSess).
Responders 
are not re-randomized at any stage.  As stated in the Introduction, there are three primary outcomes of interest.  In
order of priority, these are:  (i) risky behavior (binary), measured as condomless anal 
sex (CAS) with a non-serious partner in the past three months; (ii) HIV/STI testing in the
past three months if sexually active (binary); and (iii) subject-reported self-efficacy (continuous).  
All outcomes are coded so that higher is better, e.g., CAS is coded as an indicator
of the absence of CAS.  Self-efficacy is  re-scaled to 
take values in $[0,1]$.  The set of clinically important cutoffs is 
taken as $\bdelta = (\delta_1, \delta_2, \delta_3) = (0.05, 0.05, 0.05)$.

At each stage, we estimated the Q-functions using random forests
\citep[][]{breiman2001random} as implemented in R under the
default settings (cran.r-project.org/web/packages/randomForest/).  
We adjusted for baseline patient demographic, socio-economic, and behavioral
variables as well as interim outcomes.  See Table \ref{tailoringVariables} for a list of 
variables.  After removing
subjects who did not begin their initial treatment, 
our analyses are based on $n=2m=1024$ subjects.  Multiple imputation was used
for both item missingness and dropout following the 
guidelines given in \citet[][]{shortreed2014multiple}.  All inference
is based on sample splitting as described in Section 4.

\begin{table}[h!]
\begin{tabular}{lp{7cm}l}
Variable name & Description & Type \\ \hline
Spanish &  Spanish speaker & Binary \\ 
Latinx &   Self-identified as Latinx & Binary \\
American Indian or Alaskan Native & Self-identified as American Indian or Alaskan Native & Binary\\
Asian & Self-identified as Asian & Binary \\
Black & Self-identified as Black & Binary \\
Native Hawaiian or PI & Self-identified as Native Hawaiian or Pacific Islander & Binary \\ 
White & Self-identified as White & Binary \\ 
Other race & Self-identified as Other race & Binary \\
Age        & Age in years time of intervention & Integer \\  
Urban      & Resides in urban area & Binary \\
SES 1       & Receive public assistance & Binary \\
SES 2       & Perceived family affluence & Ordinal \\ 
AS          & Engaged in anal sex in past 3 months & Binary \\
CAS         & Engaged in condomless anal sex in past 3 months & Binary \\ 
CAS casual   & Engaged in condomless anal sex with non-serious partner in past 3 months & Binary \\ 
Num partners & Number of sexual partners in past 3 months & Integer \\ 
Num AS acts & Number of AS acts in past 3 months & Integer \\ 
NUM CAS acts & Number of CAS acts in past 3 months & Integer \\ 
Non-pres drugs & Number of non-prescription drugs used past 3 months & Integer \\ 
Alcohol        & Alcohol quantity-frequency score & Integer \\ 
Perceived vulnerability & HIV/STI vulnerability score & Integer \\ 
HIV attitude & Attitude toward HIV prevention & Integer \\ 
HIV social norms & Perceived social/behavioral norms toward HIV among peers & Integer \\ 
HIV informed & Knowledge of HIV & Integer \\ 
HIV behavioral self-efficacy & Self-efficacy for HIV prevention & Integer \\ 
HIV safe sex self-efficacy  & Self-efficacy for engaging in safe sex & Integer \\ 
STI test & Tested for STI in past 3 months & Binary \\
HIV test & Test for HIV in past 3 months & Binary \\ 
Condom intention & Intention to use condoms & Continuous \\
Condom self-efficacy & Self-efficacy towards condom use & Continuous 
\end{tabular}
\caption{\label{tailoringVariables} Tailoring variables used in the estimation
of an optimal treatment strategy. Each variable is measured at each stage preceding 
treatment assignment and the most up-to-date measurements are used as inputs to
the decision rules.}
\end{table}

Our class of decision rules at the second stage was constructed in the same
way as in our simulation experiments. Define
\begin{equation*}
\bPi^2 = \left\lbrace \pi^2(\bh^2;\bLambda) 
= \arg\max_{a^2\in\psi^2(\bh^2)}\sum_{\ell=1}^L\lambda_\ell Q_\ell^2(\bh^2, a^2
\,:\, \blambda \in [0,1]^L \mbox{ and } \sum_{\ell=1}^L\lambda_\ell = 1
\right\rbrace,
\end{equation*}
so that $\bPi$ is the set of optimal decision rules for convex combinations 
of the outcomes.  Let
$\widehat{\bPi}_m^2$ be the resulting class of rules obtained by replacing the 
$Q$-functions with
their estimated counterparts and by taking a large sample of $\blambda$ values
from the $L$-dim probability simplex (we used 100,000 samples).  Again, as in the
simulation experiments, the
set of first stage decision rules is unrestricted save for the restriction
that $\pi^1(\bh^1) \in \psi^1(\bh^1)$.  Additional description along with code to replicate
our estimators is provided in the Supplemental Materials.

To form a baseline for comparison, we estimate  
optimal regimes for each outcome separately, i.e., 
$\bpi_{\ell}^{\mathrm{opt}} =\arg\max_{\bpi\in\bPi}V_{\ell}(\bpi)$, using
Q-learning with random forests.  We also evaluate the four fixed regimes
which are embedded (after the burn-in) in the trial: (E1) assign SSE 2.0+ initially 
and subsequently assign SS to non-responders; (E2) assign SSE 2.0+ initially and subsequently SESS to non-responders;
(E3) assign SS+ initially and subsequently SSB to non-responders; and (E4) assign SS+ initially
and SESS subsequently to non-responders.  

Table  \ref{studyResults} shows the estimated
values and corresponding 95\% confidence intervals
across all regimes under consideration.  It can be 
seen that the estimated prioritized regime  
dominates all other regimes in the sense that,
relative to a competing regime, it 
is never significantly worse on any outcome while
be significantly better on at least one outcome.  
Notably, the prioritized regime (as well as the regime
optimized via $Q$-learning for CAS) yields significant improvement on the rates of CAS with a non-serious
partner relative to any fixed embedded regime.  
This reduction, if realized at scale, has the
potential for a significant positive impact
on public health.  

\begin{figure}
\begin{center}
\includegraphics[clip=true, trim=0in 14in 0in 14in, width=7in]{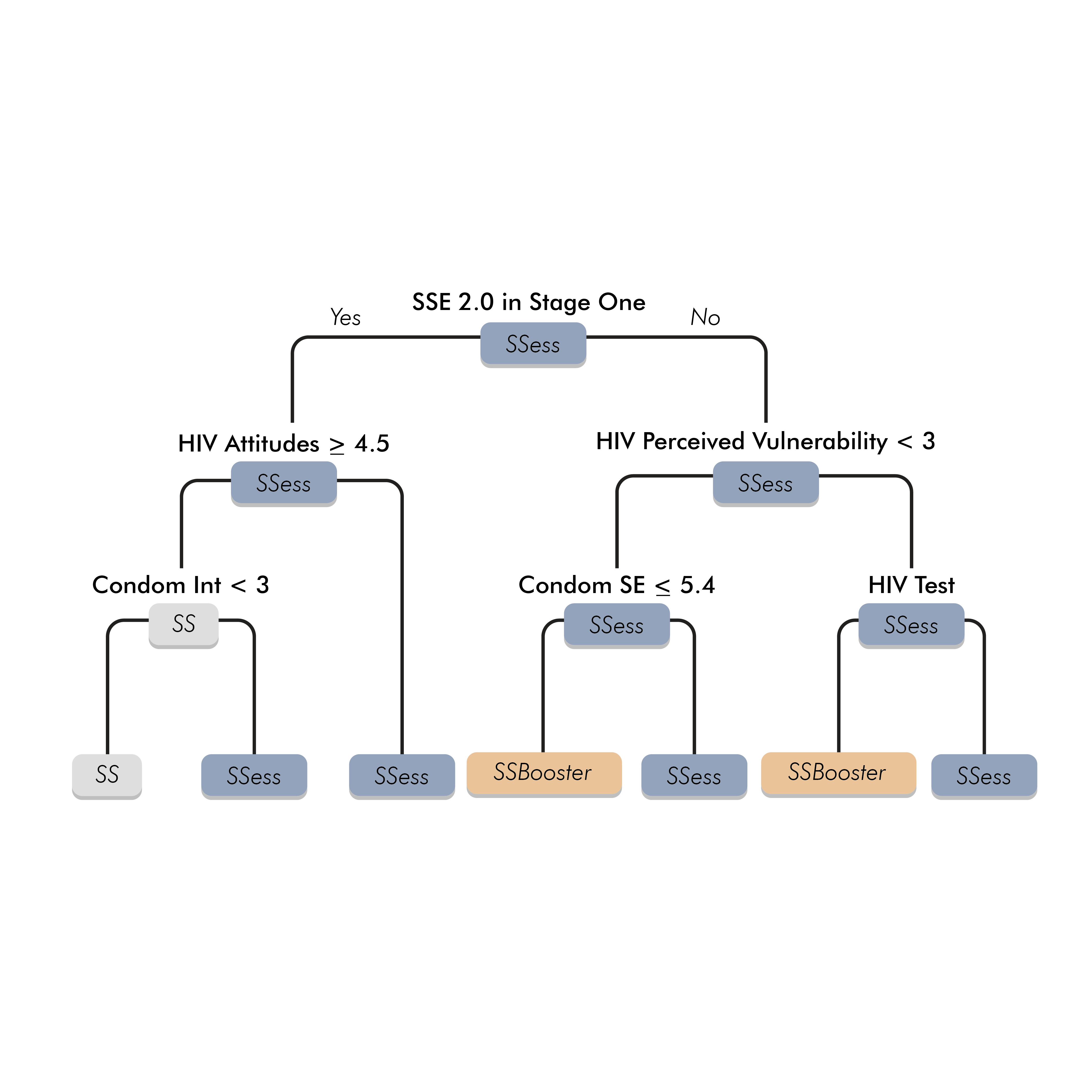}
\end{center}
\caption{\label{secondStageOptRules}
Estimated second stage decision
rule for HIV/STI prevention trial projected onto
a decision tree.  Among subjects who started on 
SSE 2.0 and did not respond, the estimated optimal rule recommends
SSess to all subjects except those with high HIV attitude
scores but low condom intention.  Among subjects who started 
on SS+ and did not respond, the estimated optimal rule
recommends SSess to subjects with low perceived HIV vulnerability 
but high condom self-efficacy or those without high perceived 
HIV vulnerability but who are not currently testing; all others 
are recommended to SSBooster.  
}
\end{figure}

To better understand the estimated prioritized
regime, we projected the decision rules
at each stage onto decision trees.  The
second stage decision rules are displayed in 
Figure \ref{secondStageOptRules}.  Among
subjects who started on SS+ and did not respond, 
the estimated optimal
second stage rule recommends  SESS to those who
are not engaging in AS and who have high levels
of knowledge, self-efficacy, or intentions and
who do not report using more than two non-prescription 
drugs in the past three months; others starting
on SS+ are recommended to SSB.  Among subjects who started
on SSE 2.0+ and did not respond, the estimated optimal
second stage rule recommends SESS to those who again
have high intentions or self-efficacy and who
are older, do not have a high alcohol quantity-frequency
score, or did not report more than 5 CAS acts; others starting
on SSE 2.0+ are recommended to SS.  Thus, the second
stage rule appears to recommend SESS to those who are 
at lower risk for STI/HIVs, i.e., higher levels of knowledge, intentions, self-efficacy along with lower sexual
activity and substance use.  

The estimated optimal first stage rule recommends SS+ to 
  those subjects are not testing, are younger, 
  do not have strong social norms around HIV prevention,
  and are not AI/AN as wells as those who are older, 
  are engaging in AS, and have high HIV self-efficacy.
  All others are recommended to SSE 2.0+.  

\begin{table}[h!]
\begin{tabular}{l|rr|rr|rr}
Regime & $V_1$ & 95\% CI & $V_2$ & 95\% CI & $V_3$ & 
95\% CI  \\ \hline
Prioritized & $0.91$ & $[0.89, 0.93]$ & $0.09$ & $[0.07, 0.11]$ & $0.79$ & $[0.78, 0.79]$\\[3pt]
$Q$-learning for $Y_1$ & $0.91$ & $[0.89, 0.94]$ & $0.04$ & $[0.03, 0.06]$ &
$0.73$ & $[0.72, 0.75]$ \\[3pt] 
$Q$-learning for $Y_2$ & $0.78$ & $[0.76, 0.81]$ & $0.13$ & $[0.11, 0.16]$
& $0.73$ & $[0.71, 0.74]$ \\[3pt] 
$Q$-learning for $Y_3$ & $0.80$  & $[0.78, 0.83]$ & $0.08$ & $[0.06, 0.09]$ & $0.80$ 
& $[0.79, 0.81]$\\[3pt]
Embedded regime (E1) & $0.83$ & $[0.80, 0.85]$ & $0.05$ &$[0.04, 0.06]$ & $0.72$ & $[0.71, 0.73]$
\\[3pt]
Embedded regime (E2) & $0.79$ & $[0.76, 0.82]$ & $0.09$ & $[0.07, 0.11]$ & $0.74$ & $[0.73, 0.76]$
\\[3pt]
Embedded regime (E3) & $0.80$ & $[0.77, 0.83]$ & $0.10$ & $[0.09, 0.12]$ & 
$0.74$ & $[0.73, 0.75]$ \\[3pt]
Embedded regime (E4) & $0.82$ & $[0.80, 0.85]$ & $0.08$ & $[0.06, 0.10]$
 & $0.74$ & $[0.73, 0.75]$ \end{tabular}
 \caption{\label{studyResults} Estimated mean outcomes for 
 CAS ($V_1$), HIV testing ($V_2$), and condom 
 self-efficacy ($V_3$) under the prioritized 
 outcome framework, $Q$-learning, and for the 
 embedded regimes (E1)-(E4). It can be seen that the
 prioritized regime dominates all the other regimes in
 the sense that it is never significantly worse on
 any outcome while being significantly better on 
 at least one outcome.  
}
\end{table}


\begin{figure}
\begin{center}
\includegraphics[clip=true, trim=0in 14in 0in 14in, width=7in]{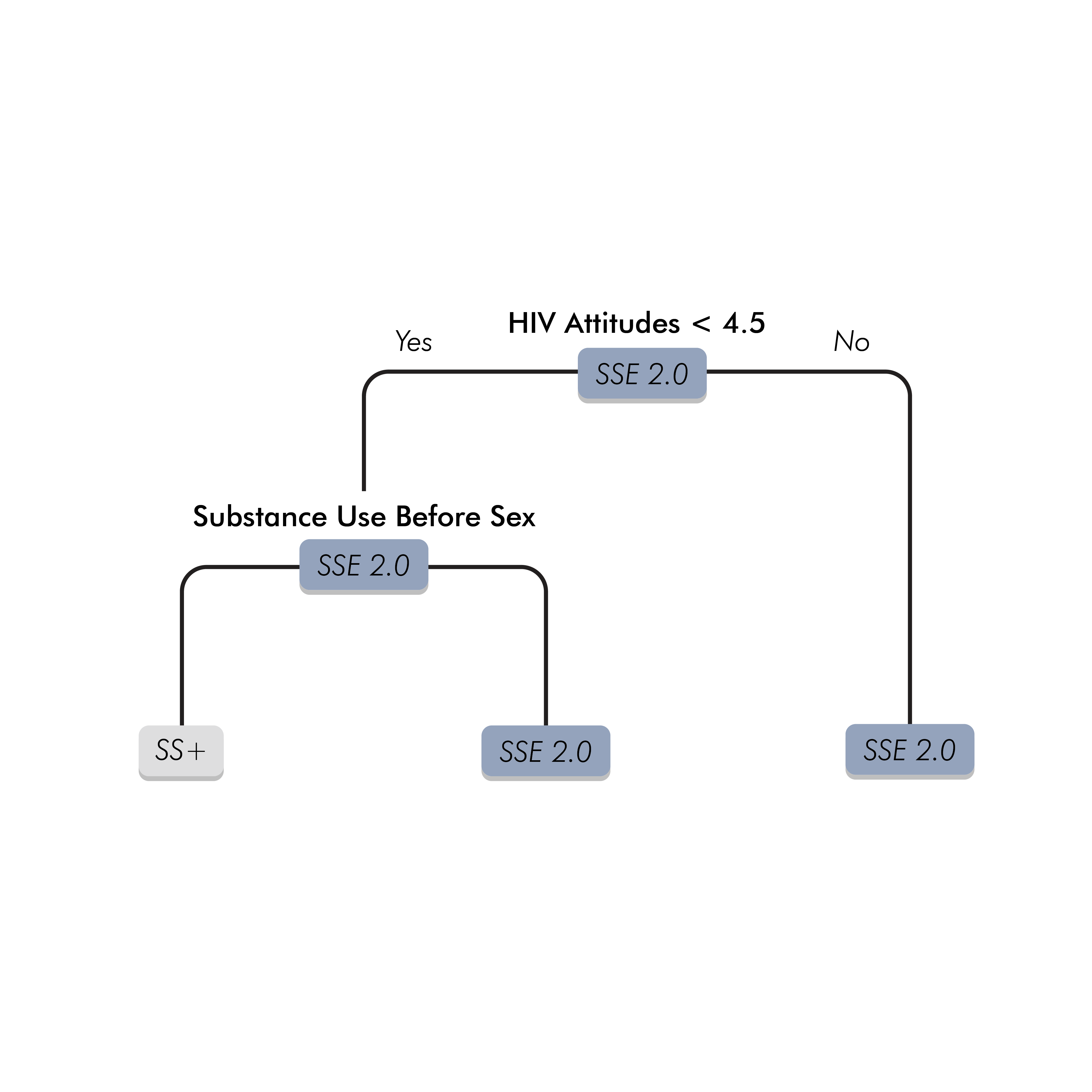}
\end{center}
\caption{
  \label{firstStageOptRule} Estimated optimal first stage
  decision rule for HIV/STI prevention trial projected
  onto a decision tree.  The rule recommends SS to patients
  with low HIV attitudes and who report substance use before sex,
  all others are recommended to SSE 2.0.    
}
\end{figure}




\section{Discussion}\label{sec:Discussion}  
We proposed a definition of an optimal treatment
regime when there are multiple prioritized 
outcomes.  Using inverse reinforcement learning,
we defined an optimal composite
outcome within a pre-specified class 
as the one yielding highest utility under
the optimal treatment regime.  We proposed
estimators of the optimal treatment regime 
and optimal composite outcome and derived
large sample inference methods.  

There are number of interesting directions in which
this work might be extended.  One is to combine the
current work with the 
ideas of a set-valued regime 
\citep[][]{laber2014set,wu2016set,ertefaie2016identifying,lizotte2016multi}.
Thus, rather than identifying a single regime one could enumerate
all regimes that are acceptable given prioritized outcomes
and clinically meaningful differences.  An advantage of this
approach is that it allows any `tie-breaking' between treatments
to depend on factors not captured by the model, e.g., personal
preference, resource constraints, etc.   Another direction for
future work is the extension of prioritized outcomes to 
the infinite horizon setting.  With an indefinite treatment
horizon, one can adaptively shape a composite utility function to drive 
behavior change in accordance with a theoretical behavior model;
e.g., a treatment strategy might first target 
self-efficacy and perceived 
vulnerability to STI/HIV infection and, only after these outcomes 
are within some targeted
range, does the treatment regime focus on testing behavior.

\bibliographystyle{Chicago}
\bibliography{nw}

\begin{thebibliography}{}

\bibitem[\protect\citeauthoryear{Arora and Doshi}{Arora and
  Doshi}{2021}]{arora2021survey}
Arora, S. and P.~Doshi (2021).
\newblock A survey of inverse reinforcement learning: Challenges, methods and
  progress.
\newblock {\em Artificial Intelligence\/}~{\em 297}, 103500.

\bibitem[\protect\citeauthoryear{Breiman}{Breiman}{2001}]{breiman2001random}
Breiman, L. (2001).
\newblock Random forests.
\newblock {\em Machine learning\/}~{\em 45\/}(1), 5--32.

\bibitem[\protect\citeauthoryear{Butler, Laber, Davis, and Kosorok}{Butler
  et~al.}{2018}]{butler2018incorporating}
Butler, E.~L., E.~B. Laber, S.~M. Davis, and M.~R. Kosorok (2018).
\newblock Incorporating patient preferences into estimation of optimal
  individualized treatment rules.
\newblock {\em Biometrics\/}~{\em 74\/}(1), 18--26.

\bibitem[\protect\citeauthoryear{Buyse}{Buyse}{2010}]{buyse2010generalized}
Buyse, M. (2010).
\newblock Generalized pairwise comparisons of prioritized outcomes in the
  two-sample problem.
\newblock {\em Statistics in medicine\/}~{\em 29\/}(30), 3245--3257.

\bibitem[\protect\citeauthoryear{Embretson and Reise}{Embretson and
  Reise}{2013}]{embretson2013item}
Embretson, S.~E. and S.~P. Reise (2013).
\newblock {\em Item response theory}.
\newblock Psychology Press.

\bibitem[\protect\citeauthoryear{Ertefaie, McKay, Oslin, and
  Strawderman}{Ertefaie et~al.}{2021}]{ertefaie2021robust}
Ertefaie, A., J.~R. McKay, D.~Oslin, and R.~L. Strawderman (2021).
\newblock Robust q-learning.
\newblock {\em Journal of the American Statistical Association\/}~{\em
  116\/}(533), 368--381.

\bibitem[\protect\citeauthoryear{Ertefaie, Wu, Lynch, and Nahum-Shani}{Ertefaie
  et~al.}{2016}]{ertefaie2016identifying}
Ertefaie, A., T.~Wu, K.~G. Lynch, and I.~Nahum-Shani (2016).
\newblock Identifying a set that contains the best dynamic treatment regimes.
\newblock {\em Biostatistics\/}~{\em 17\/}(1), 135--148.

\bibitem[\protect\citeauthoryear{Finkelstein and Schoenfeld}{Finkelstein and
  Schoenfeld}{1999}]{finkelstein1999combining}
Finkelstein, D.~M. and D.~A. Schoenfeld (1999).
\newblock Combining mortality and longitudinal measures in clinical trials.
\newblock {\em Statistics in medicine\/}~{\em 18\/}(11), 1341--1354.

\bibitem[\protect\citeauthoryear{Hadfield-Menell, Milli, Abbeel, Russell, and
  Dragan}{Hadfield-Menell et~al.}{2017}]{hadfield2017inverse}
Hadfield-Menell, D., S.~Milli, P.~Abbeel, S.~J. Russell, and A.~Dragan (2017).
\newblock Inverse reward design.
\newblock {\em Advances in neural information processing systems\/}~{\em 30}.

\bibitem[\protect\citeauthoryear{Illenberger, Spieker, and Mitra}{Illenberger
  et~al.}{2021}]{illenberger2021identifying}
Illenberger, N., A.~J. Spieker, and N.~Mitra (2021).
\newblock Identifying optimally cost-effective dynamic treatment regimes with a
  q-learning approach.
\newblock {\em arXiv preprint arXiv:2107.03441\/}.

\bibitem[\protect\citeauthoryear{Laber, Lizotte, and Ferguson}{Laber
  et~al.}{2014}]{laber2014set}
Laber, E.~B., D.~J. Lizotte, and B.~Ferguson (2014).
\newblock Set-valued dynamic treatment regimes for competing outcomes.
\newblock {\em Biometrics\/}.

\bibitem[\protect\citeauthoryear{Laber, Lizotte, Qian, Pelham, and
  Murphy}{Laber et~al.}{2014}]{laber2014statistical}
Laber, E.~B., D.~J. Lizotte, M.~Qian, W.~E. Pelham, and S.~A. Murphy (2014).
\newblock Dynamic treatment regimes: Technical challenges and applications.
\newblock {\em Electronic journal of statistics\/}~{\em 8\/}(1), 1225--1272.

\bibitem[\protect\citeauthoryear{Laber, Wu, Munera, Lipkovich, Colucci, and
  Ripa}{Laber et~al.}{2018}]{laber2018identifying}
Laber, E.~B., F.~Wu, C.~Munera, I.~Lipkovich, S.~Colucci, and S.~Ripa (2018).
\newblock Identifying optimal dosage regimes under safety constraints: An
  application to long term opioid treatment of chronic pain.
\newblock {\em Statistics in medicine\/}~{\em 37\/}(9), 1407--1418.

\bibitem[\protect\citeauthoryear{Lavori and Dawson}{Lavori and
  Dawson}{2000}]{lavori2000design}
Lavori, P. and R.~Dawson (2000).
\newblock A design for testing clinical strategies: biased adaptive
  within-subject randomization.
\newblock {\em Journal of the Royal Statistical Society: Series A (Statistics
  in Society)\/}~{\em 163\/}(1), 29--38.

\bibitem[\protect\citeauthoryear{Lavori and Dawson}{Lavori and
  Dawson}{2004}]{lavori2004dynamic}
Lavori, P.~W. and R.~Dawson (2004).
\newblock Dynamic treatment regimes: practical design considerations.
\newblock {\em Clinical trials\/}~{\em 1\/}(1), 9--20.

\bibitem[\protect\citeauthoryear{Liang and Yu}{Liang and
  Yu}{2020}]{liang2020relative}
Liang, M. and M.~Yu (2020).
\newblock Relative contrast estimation and inference for treatment
  recommendation.
\newblock {\em arXiv preprint arXiv:2010.13904\/}.

\bibitem[\protect\citeauthoryear{Linn, Laber, and Stefanski}{Linn
  et~al.}{2015}]{linn2015chapter}
Linn, K.~A., E.~B. Laber, and L.~A. Stefanski (2015).
\newblock Chapter 15: Estimation of dynamic treatment regimes for complex
  outcomes: balancing benefits and risks.
\newblock In {\em Adaptive treatment strategies in practice: Planning trials
  and analyzing data for personalized medicine}, pp.\  249--262. SIAM.

\bibitem[\protect\citeauthoryear{Lizotte, Bowling, and Murphy}{Lizotte
  et~al.}{2012}]{lizotte2012linear}
Lizotte, D.~J., M.~Bowling, and S.~A. Murphy (2012).
\newblock Linear fitted-q iteration with multiple reward functions.
\newblock {\em The Journal of Machine Learning Research\/}~{\em 13\/}(1),
  3253--3295.

\bibitem[\protect\citeauthoryear{Lizotte and Laber}{Lizotte and
  Laber}{2016}]{lizotte2016multi}
Lizotte, D.~J. and E.~B. Laber (2016).
\newblock Multi-objective markov decision processes for data-driven decision
  support.
\newblock {\em The Journal of Machine Learning Research\/}~{\em 17\/}(1),
  7378--7405.

\bibitem[\protect\citeauthoryear{Luedtke and Van Der~Laan}{Luedtke and Van
  Der~Laan}{2016}]{luedtke2016statistical}
Luedtke, A.~R. and M.~J. Van Der~Laan (2016).
\newblock Statistical inference for the mean outcome under a possibly
  non-unique optimal treatment strategy.
\newblock {\em Annals of statistics\/}~{\em 44\/}(2), 713.

\bibitem[\protect\citeauthoryear{Mao and Wang}{Mao and
  Wang}{2021}]{mao2021class}
Mao, L. and T.~Wang (2021).
\newblock A class of proportional win-fractions regression models for composite
  outcomes.
\newblock {\em Biometrics\/}~{\em 77\/}(4), 1265--1275.

\bibitem[\protect\citeauthoryear{Murphy}{Murphy}{2005}]{murphy2005experimental}
Murphy, S.~A. (2005).
\newblock An experimental design for the development of adaptive treatment
  strategies.
\newblock {\em Statistics in medicine\/}~{\em 24\/}(10), 1455--1481.

\bibitem[\protect\citeauthoryear{Mustanski, Moskowitz, Moran, Newcomb,
  Macapagal, Rodriguez-D{\'\i}az, Rendina, Laber, Li, Matson, et~al.}{Mustanski
  et~al.}{2020}]{mustanski2020evaluation}
Mustanski, B., D.~A. Moskowitz, K.~O. Moran, M.~E. Newcomb, K.~Macapagal,
  C.~Rodriguez-D{\'\i}az, H.~J. Rendina, E.~B. Laber, D.~H. Li, M.~Matson,
  et~al. (2020).
\newblock Evaluation of a stepped-care ehealth hiv prevention program for
  diverse adolescent men who have sex with men: protocol for a hybrid type 1
  effectiveness implementation trial of smart.
\newblock {\em JMIR research protocols\/}~{\em 9\/}(8), e19701.

\bibitem[\protect\citeauthoryear{Oakes}{Oakes}{2016}]{oakes2016win}
Oakes, D. (2016).
\newblock On the win-ratio statistic in clinical trials with multiple types of
  event.
\newblock {\em Biometrika\/}~{\em 103\/}(3), 742--745.

\bibitem[\protect\citeauthoryear{Pocock, Ariti, Collier, and Wang}{Pocock
  et~al.}{2012}]{pocock2012win}
Pocock, S.~J., C.~A. Ariti, T.~J. Collier, and D.~Wang (2012).
\newblock The win ratio: a new approach to the analysis of composite endpoints
  in clinical trials based on clinical priorities.
\newblock {\em European heart journal\/}~{\em 33\/}(2), 176--182.

\bibitem[\protect\citeauthoryear{Rose, Laber, Davidian, Tsiatis, Zhao, and
  Kosorok}{Rose et~al.}{2019}]{rose2019sample}
Rose, E.~J., E.~B. Laber, M.~Davidian, A.~A. Tsiatis, Y.-Q. Zhao, and M.~R.
  Kosorok (2019).
\newblock Sample size calculations for smarts.
\newblock {\em arXiv preprint arXiv:1906.06646\/}.

\bibitem[\protect\citeauthoryear{Shi, Zhang, Lu, and Song}{Shi
  et~al.}{2020}]{shi2020statistical}
Shi, C., S.~Zhang, W.~Lu, and R.~Song (2020).
\newblock Statistical inference of the value function for reinforcement
  learning in infinite horizon settings.
\newblock {\em arXiv preprint arXiv:2001.04515\/}.

\bibitem[\protect\citeauthoryear{Shortreed, Laber, Scott~Stroup, Pineau, and
  Murphy}{Shortreed et~al.}{2014}]{shortreed2014multiple}
Shortreed, S.~M., E.~Laber, T.~Scott~Stroup, J.~Pineau, and S.~A. Murphy
  (2014).
\newblock A multiple imputation strategy for sequential multiple assignment
  randomized trials.
\newblock {\em Statistics in medicine\/}~{\em 33\/}(24), 4202--4214.

\bibitem[\protect\citeauthoryear{Sundararajan and Najmi}{Sundararajan and
  Najmi}{2020}]{sundararajan2020many}
Sundararajan, M. and A.~Najmi (2020).
\newblock The many shapley values for model explanation.
\newblock In {\em International conference on machine learning}, pp.\
  9269--9278. PMLR.

\bibitem[\protect\citeauthoryear{Thall}{Thall}{2020}]{thall2020statistical}
Thall, P.~F. (2020).
\newblock {\em Statistical remedies for medical researchers}.
\newblock Springer.

\bibitem[\protect\citeauthoryear{Tsiatis, Davidian, Holloway, and
  Laber}{Tsiatis et~al.}{2020}]{tsiatis2019dynamic}
Tsiatis, A.~A., M.~Davidian, S.~T. Holloway, and E.~B. Laber (2020).
\newblock {\em Dynamic Treatment Regimes: Statistical Methods for Precision
  Medicine}.
\newblock CRC press.

\bibitem[\protect\citeauthoryear{van~der Laan and Petersen}{van~der Laan and
  Petersen}{2007}]{van2007causal}
van~der Laan, M.~J. and M.~L. Petersen (2007).
\newblock Causal effect models for realistic individualized treatment and
  intention to treat rules.
\newblock {\em International Journal of Biostatistics\/}~{\em 3\/}(1), 3.

\bibitem[\protect\citeauthoryear{Wang, Zhao, and Zheng}{Wang
  et~al.}{2020}]{wang2020learning}
Wang, Y., Y.-Q. Zhao, and Y.~Zheng (2020).
\newblock Learning-based biomarker-assisted rules for optimized clinical
  benefit under a risk constraint.
\newblock {\em Biometrics\/}~{\em 76\/}(3), 853--862.

\bibitem[\protect\citeauthoryear{Wasserman, Ramdas, and Balakrishnan}{Wasserman
  et~al.}{2020}]{wasserman2020universal}
Wasserman, L., A.~Ramdas, and S.~Balakrishnan (2020).
\newblock Universal inference.
\newblock {\em Proceedings of the National Academy of Sciences\/}~{\em
  117\/}(29), 16880--16890.

\bibitem[\protect\citeauthoryear{Wu}{Wu}{2016}]{wu2016set}
Wu, T. (2016).
\newblock {\em Set Valued Dynamic Treatment Regimes.}
\newblock Ph.\ D. thesis.

\bibitem[\protect\citeauthoryear{Zhang, Laber, Tsiatis, and Davidian}{Zhang
  et~al.}{2018}]{listBasedRegimes}
Zhang, Y., E.~Laber, A.~Tsiatis, and M.~Davidian (2018).
\newblock Interpretable dynamic treatment regimes.
\newblock {\em Journal of the American Statistical Association\/}~{\em
  113\/}(524), 1541--1549.

\bibitem[\protect\citeauthoryear{Zhao, Zeng, Laber, and Kosorok}{Zhao
  et~al.}{2015}]{yingqi2}
Zhao, Y., D.~Zeng, E.~B. Laber, and M.~R. Kosorok (2015).
\newblock New statistical learning methods for estimating optimal dynamic
  treatment regimes.
\newblock {\em Journal of the American Statistical Association\/}~{\em
  110\/}(510), 583--598.

\end{thebibliography}

\end{document}